\documentclass{ws-ijmpa}
\usepackage[super,compress]{cite}
\usepackage{graphicx}
\usepackage{color}
\usepackage{physics}
\usepackage{multirow}
\usepackage{arydshln}
\usepackage{hyperref}
\usepackage{url}

\usepackage{eso-pic}
\newcommand{\reportnum}[2]{
  \AddToShipoutPictureBG*{%
    \AtPageUpperLeft{%
      \hspace{0.75\paperwidth}%
      \raisebox{#1\baselineskip}{%
        \makebox[0pt][l]{\textnormal{#2}}
  }}}%
}
\reportnum{-7}{\hspace{17cm}HUPD-2408}

\begin{document}

\markboth{Ken-Ichi Ishikawa, Masanori Okawa, Hironori Takei}{Perturbative gradient flow coupling of ...}

%
\catchline{}{}{}{}{}
%

\title{Perturbative gradient flow coupling of the twisted Eguchi--Kawai model with the numerical stochastic perturbation theory}

\author{
        Ken-Ichi Ishikawa$^{1,2,}$\footnote{ishikawa@theo.phys.sci.hiroshima-u.ac.jp},
        Masanori Okawa$^{1,}$\footnote{okawa@hiroshima-u.ac.jp},
        Hironori Takei$^{1,}$\footnote{t-hironori@hiroshima-u.ac.jp}}

\address{
$^{1}$Graduate School of Advanced Science and Engineering, Hiroshima University, Higashi-Hiroshima, Hiroshima 739-8526, Japan\\
$^{2}$Core of Research for the Energetic Universe, Graduate School of Advanced Science and Engineering, Hiroshima University, Higashi-Hiroshima, Hiroshima 739-8526, Japan}

\maketitle


\newcommand{\MSbar}{$\overline{\mathrm{MS}}$\ }
\newcommand{\figscale}{0.40}

\begin{abstract}
{
The gradient flow scheme has emerged as a prominent nonperturbative renormalization scheme 
on the lattice, where flow time is introduced to define the renormalization scale.
In this study we perturbatively compute the gradient flow coupling 
for the SU($N$) Yang--Mills theory in the large-$N$ limit
in terms of the lattice bare coupling up to three-loop order. 
This is achieved by combining the twisted Eguchi--Kawai model with the numerical stochastic perturbation theory.
We analyze the flow time dependence of the perturbative coefficients to determine the perturbative beta function 
coefficients, successfully computing the one-loop coefficient in the large-$N$ limit using three matrix sizes $N=289,441,529$.
However, the higher-order coefficients are affected by large statistical errors.
We also explore the potential for reducing these statistical errors through variance 
reduction combined with the large-$N$ factorization property of the SU($N$) Yang-Mills theory,
and estimate the required number of samples for the precise determination of the higher-order coefficients.
}
\end{abstract}




\section{Introduction}
\label{sec:introduction}

The gradient flow scheme, originally developed for lattice quantum chromodynamics (QCD)~\cite{Narayanan_2006,L_scher_2010,L_scher_2011}, 
has emerged as an important renormalization scheme in both lattice and continuum field theories.
In QCD, nonperturbative matrix elements can be computed using the lattice QCD method.
However, matching or renormalizing these lattice QCD operators to the \MSbar scheme is an essential process 
because experimental results in theoretical QCD are typically expressed in terms of the \MSbar (or $\mathrm{MS}$) scheme.
The gradient flow scheme serves as an intermediate scheme between the \MSbar and lattice schemes,
avoiding direct matching between the two.
This is beneficial because the gradient flow scheme has desirable features,
such as regularization independence and reduced statistical fluctuations in the nonperturbative expectation values during lattice numerical simulations.
Applications of the gradient flow scheme are apparent in various studies, such as Refs.~\citen{borsanyi_2012,Zoltan_Fodor_2012,Zoltan_Fodor_2014},
with further extensions and applications documented in Refs.~\citen{Fritzsch_2013,Bonati_2014,Shindler_2014,Monahan_2014}. 

Previous studies have assessed the matching between the gradient flow coupling and the \MSbar coupling in QCD using dimensional regularization 
in continuum theory\cite{Harlander_2016,Artz_2019,Lange_2021}.
While nonperturbative renormalization in lattice QCD is a powerful method,
the perturbative renormalization plays a notable role in providing information regarding theoretical aspects, such as the effects of lattice regularization and finite volume corrections.
While perturbative calculations within the gradient flow scheme on the lattice can be accomplished using the standard Feynman diagram method,
progressing beyond three-loop orders become challenging owing to the complexities introduced by lattice regularization and gradient flow evolution.

A promising approach for numerically evaluating perturbative expansion coefficients in lattice field theory 
is the numerical stochastic perturbation theory (NSPT),
which allows automatic computations of these perturbative coefficients without Feynman diagrams.
NSPT is based on the perturbative expansion of the Langevin equation in the stochastic quantization\cite{Parisi_Wu,Floratos_1983}.
The first studies on NSPT employed the Langevin-type Monte Carlo algorithm\cite{Renzo_1993,Renzo_1994,Renzo_1994_2,Renzo_1995},
with subsequent attempts using algorithms based on Kramers, and hybrid molecular dynamics (HMD)\cite{brida_2017,Brida_Luscher_SMD_NSPT}.

Using NSPT, the gradient flow coupling can be evaluated in terms of the lattice bare coupling\cite{brida2013_GF,brida2015_GF_phi4,brida_2016,Brida_Luscher_SMD_NSPT}. 
By combining the NSPT results with the continuum relation between the \MSbar coupling and the gradient flow coupling, 
we can derive the \MSbar coupling in terms of the lattice bare coupling.
In SU($N$) gauge theory, the matching between the \MSbar coupling and the gradient flow coupling
is computed at the two-loop level, along with the first non-universal beta function coefficient,
using dimensional regularization\cite{Harlander_2016,Artz_2019,Lange_2021}.
Once a higher-order relation between the \MSbar coupling and the gradient flow coupling is derived, 
the \MSbar coupling can be obtained in temrs of the lattice bare coupling at the relevant order by performing 
higher-order calculations of the gradient flow coupling using the NSPT method.
This approach can also be extended to other renormalization constants between the \MSbar and gradient flow schemes,
making it worthwhile to compute higher-order perturbation of the gradient flow scheme using NSPT.

This study focuses on the gradient flow coupling within the infinite volume and large-$N$ limits of the SU($N$) Yang--Mills theory,
assessing the application feasibility of NSPT in the gradient flow scheme.
Previous studies\cite{Renzo_2001,Horsley_2012,Bauer_2012,Brambilla_2013,Bauer_2013} have explored the application of NSPT to
higher-order calculations in lattice QCD (SU($3$)), investigating effects such as renormalon and the gluon condensate.
However, the difficulties arises in finite volume corrections and the need for careful considerations of the infinite volume
limit based on renormalization group properties.
To simplify the treatment of finite volume corrections in the feasibility study of the gradient flow with the NSPT, 
we rely on the large-$N$ limit of the SU($N$) Yang--Mills theory,
where volume dependence disappears in the large-$N$ limit.

The large-$N$ limit for the gauge theory, first proposed by Gerard~'t~Hooft\cite{HOOFT_1974},
simplifies the theory by ensuring that only certain diagrams including gluon loops, 
referred to as planar diagrams, contribute to the dynamics.
Despite this simplification, the large-$N$ limit retains key features of the gauge theory,
making it an important tool for both theoretical understanding and the study of real QCD,
where QCD can be approached through $1/N^2$ corrections\cite{Donini_2016,Thomas_2016,Lucini_2004}.
Furthermore, the large-$N$ limit simplifies finite volume dependence,
as mentioned earlier, owing to the large-$N$ factorization property\cite{Witten_1979,makeenko_2000_largen,Lucini_2013,Muguel_largeN_2018}.
This property offers insights into the equivalence between the SU($N$) Yang--Mills theory and the matrix model proposed 
by Eguchi and Kawai\cite{Eguchi_Kawai} in the large-$N$ limit.
Gonz\'{a}lez-Arroyo and Okawa improved the Eguchi--Kawai model by introducing the twisted boundary condition,
leading to the twisted Eguchi--Kawai (TEK) model\cite{TEK_1983_lat,TEK_1983}.
Previous studies have investigated nonperturbatively the TEK model and validated its 
equivalence to the SU($N$) Yang--Mills theory in the large-$N$ limit\cite{Gonz_lez_Arroyo_2010,arroyo2014_test_tek}.

The twisted boundary condition plays a crucial role in defining the gradient flow coupling renormalized 
at a finite volume length, referred to as the twisted gradient flow scheme\cite{ramos_2013_tgf,ramos_2014_tgf,perez2014_tgf,golan2021_tgf,kishika2017_lambda}.
This scheme is for lattice simulations, facilitating the definition of the step scaling evolution of
the renormalized coupling\cite{cheng2014_ssf,luscher2014_ssf,perez2014suinfty}.
The one-loop calculation of the twisted gradient flow coupling has been done with the continuum perturbation theory\cite{margarita_2018,margarita_2019}.
In this study, we explore the applicability of the NSPT to the gradient flow scheme
in the large-$N$ limit of the SU($N$) Yang--Mills theory using the TEK model.

We compute the gradient flow coupling up to the three-loop order using
the NSPT for the TEK model in the large-$N$ limit as follows:
\begin{align}
    \lambda_\rho(\mu) &= \lambda_0 + r_1(\hat{t}) \lambda_0^2 + r_2(\hat{t}) \lambda_0^3 + r_3(\hat{t})\lambda_0^4 + \cdots,
    \label{eq:rho_bare}
\end{align}
where $\lambda_{\rho}$ and $\lambda_0$ denote the gradient flow coupling and lattice bare coupling, respectively.
The renormalization scale $\mu$, dimensionless flow time $\hat{t}=t/a^2$, and flow time parameter $\rho$ 
are related as $\mu^2 t=\rho$, where $a$ is the lattice cut-off.
The first two coefficients, $r_1(\hat{t})$ and $r_2(\hat{t})$, are derived 
by combining the following analytic relations:
\begin{align}
   \lambda_{\rho}(\mu) &= \lambda_S(\mu)  + \bar{e}_1\lambda_S(\mu)^2 + \bar{e}_2\lambda_S(\mu)^3 + \cdots,
\label{eq:rho_msbar}\\
	\lambda_S(\mu) &= \lambda_0 + c_1(\mu a) \lambda_0^2 + c_2(\mu a) \lambda_0^3 + \cdots.
\label{eq:msbar_bare_lat}
\end{align} 
Previous studies\cite{Harlander_2016,Artz_2019} have derived the two-loop relation in Eq.~\eqref{eq:rho_msbar} using the continuum perturbation theory 
at finite-$N$, which can be converted to the 't~Hooft coupling form in the large-$N$ limit.
The analytical coefficients $\bar{e}_1$ and $\bar{e}_2$ are presented in the main text.
Equation~\eqref{eq:msbar_bare_lat} is derived from the results of the standard SU($N$) Wilson gauge action\cite{L_scher_1995,L_scher_2010}.
We investigate the flow time dependence, finite volume $\order{1/N^2}$ corrections,
and lattice cut-off effects for the perturbation coefficients $r_i(\hat{t})$, 
and discuss the applicability of NSPT in computing the higher-order perturbative beta function within the current setup.
Preliminary results have been presented at the 41st Lattice Conference (Lattice 2024)\cite{lattice_2024} and 
the 79th Annual Meeting of the Physical Society of Japan\cite{79th_JPS}.

The remainder of this paper is organized as follows.
In Section~\ref{sec:nspt_tek}, we briefly introduce the TEK model and the NSPT. 
Section~\ref{sec:gf_nspt} defines the gradient flow coupling in continuum and 
its NSPT version on the lattice, explaining the known analytic perturbation coefficients. 
Section~\ref{sec:numerical_result} presents the numerical results for 
the perturbation coefficients of the gradient flow coupling evaluated using the NSPT,
focusing on the flow time dependence and the large-$N$ limit.
We successfully compute the one-loop coefficient, $r_1(\hat{t})$, yielding the universal one-loop beta function;
 however, our results for the two- and three-loop coefficients were less precise owing to
large statistical errors.
For future prospects to obtain the precise higher order coefficients we discuss the sample size considering the relation between statistical error 
and variance of the perturbation coefficients based on the large-$N$ factorization property.
The paper concludes with an outlook on future work in Section~\ref{sec:summary_and_outlook}.

\section{NSPT for the TEK model}
\label{sec:nspt_tek}
This section  provides an overview of TEK model and NSPT based on the HMD method\cite{Brida_Luscher_SMD_NSPT}.

\subsection{TEK model}
\label{subsec:tek}
The TEK model is a matrix model on a one-site lattice with twisted boundary conditions.
Its partition function is expressed as 
\begin{align}
    Z =& \int\prod_{\mu=1}^{4} dU_\mu \exp\qty[-S[U]] ,    \\
    S[U] =& Nb\sum_{\mu,\nu=1}^{4} \Tr\qty[I - z_{\mu\nu}U_\mu U_\nu U_\mu^\dag U_\nu^\dag],
    \label{eq:tek_action}
\end{align}
where $U_\mu$ represent SU($N$) matrices, and $b$ is defined as 
$b \equiv {1}/{(Ng_0^2)} = {1}/{\lambda_0}$, where $\lambda_0$ is the 't~Hooft lattice bare coupling constant.
The twist factor $z_{\mu\nu}$ is defined as 
\begin{align}
    z_{\mu\nu} =& \exp\qty[\frac{2\pi i k}{\sqrt{N}}\epsilon_{\mu\nu}],
    \label{eq:twist_factor}
    \\
    \epsilon_{\mu\nu} =&
    \left\{
    \begin{array}{rl}
    +1 & \ (\mu < \nu) \\
    0  & \ (\mu = \nu) \\
    -1 & \ (\mu > \nu)
    \end{array}
    \right.,
\end{align}
where $k$ is an integer coprime with $\hat{L}=\sqrt{N}$.
The effective volume is proportional to the square of the matrix rank as $V=(a\hat{L})^4=a^4N^2$.
In the large-$N$ limit, the expectation value of the trace of a closed-loop operator for the TEK model is 
consistent with that for the SU($N$) Yang--Mills theory in the infinite volume limit,
provided that the parameter $k$\cite{Gonz_lez_Arroyo_2010} is chosen appropriately.
The finite volume effect in the TEK model manifests as finite $N$ corrections and is controlled by the phase parameter 
$\theta\equiv 2\pi|\bar{k}|/\sqrt{N}$, where $\bar{k}$ satisfies $k\bar{k}=1(\mathrm{mod}\ \sqrt{N})$.
Previous studies have demonstrated that the condition $\theta>1/9$ is sufficient for nonperturbative simulations\cite{Gonz_lez_Arroyo_2010}.
To ensure a smooth large-$N$ limit, we have to keep the phase parameter~$\theta$ at constant both nonperturbatively and perturbatively.

\subsection{NSPT}
\label{subsec:nspt}
NSPT is a powerful tool for computing higher-order coefficients in a perturbation series without Feynman diagrams.
It is implemented by expanding the field variables of a target system in terms of the coupling constant and transforming
the ordinary stochastic differential equation (the Langevin equation) into hierarchical stochastic differential equations\cite{Renzo_1993,Renzo_1994,Renzo_1994_2,Renzo_1995}.
Previous studies\cite{Brida_Luscher_SMD_NSPT,brida_2017} have extended the Langevin NSPT to use HMD and Kramers equations.

In this study, we employ the HMD-based NSPT algorithm, which has been applied to the TEK model in Ref.~\citen{Gonz_lez_Arroyo_2019}.
Below we briefly explain the NSPT simulation.
An observable $O[U_\mu]$ is evaluated nonperturbatively as the stochastic average:
\begin{align}
    \expval{O[U_\mu]} \simeq&
    \frac{1}{N_{\mathrm{sample}}} \sum_{i=1}^{N_{\mathrm{sample}}} O[U_{\mu,i}],
\end{align}
where $\{U_{\mu,i}|\ i=1,\cdots,N_{\mathrm{sample}}\}$ is a stochastic ensemble generated
using the HMD algorithm and satisfies the probability density $\prod_\mu dU_\mu\exp\qty[-S[U]]$.

To implement the NSPT, the field variable $U_\mu$ is expanded in terms of the 't~Hooft coupling $b^{-1}=\lambda_0$ as
\begin{align}
    U_\mu =& \sum_{k=0}^{\infty} \lambda_0^{k/2} U_\mu^{(k)},
    \label{eq:series_umu}
\end{align}
where $U_\mu^{(0)}=\Gamma_\mu$ represents the classical vacuum of the action Eq.~\eqref{eq:tek_action} 
and $\Gamma_\mu$ denotes the ``twist eater" matrix, satisfying the following commutation relation:
\begin{align}
    \Gamma_\mu \Gamma_\nu = z_{\nu\mu} \Gamma_\nu \Gamma_\mu,
\end{align}
where the twist factor $z_{\mu\nu}$ is defined in Eq.~\eqref{eq:twist_factor}.
The expectation value of an observable in the NSPT is evaluated as follows: 
\begin{align}
    \expval{O[U_\mu]}
    = \sum_{k=0}^{\infty} \lambda_0^{k/2} \expval{O^{(k)}[U_\mu^{(0)},\dots,U_\mu^{(k)}]},
    \label{eq:series_obs}
\end{align}
\begin{align}
    \expval{O^{(k)}[U_\mu^{(0)},\dots,U_\mu^{(k)}]} \simeq \frac{1}{N_\mathrm{sample}} \sum_{i=1}^{N_\mathrm{sample}} O^{(k)}[U_{\mu,i}^{(0)},\dots,U_{\mu,i}^{(k)}],
\end{align}
where the stochastic ensemble $\{(U_{\mu,i}^{(0)},\dots,U_{\mu,i}^{(k)},\dots)|\ i=1,\cdots,N_{\mathrm{sample}}\}$
is generated using the Langevin or HMD-based stochastic  algorithms\cite{Brida_Luscher_SMD_NSPT}.
Additional details regarding the HMD algorithm for the TEK model can be found in Ref.~\citen{Gonz_lez_Arroyo_2019}.
The coefficient $O^{(k)}[U_{\mu}^{(0)},\dots,U_{\mu}^{(k)}]$ denotes 
the $k$-th order coefficient of $O[U_\mu]$, obtained after substituting Eq.~\eqref{eq:series_umu} into $O[U_\mu]$.
The series Eq.~\eqref{eq:series_umu} can be truncated at any finite order 
for practical computations, as the $k$-th order coefficient depends only on the coefficients $\{U_\mu^{(0)},\dots,U_\mu^{(k)}\}$.
The next section focuses on the evaluation of the coefficients of the gradient flow coupling in the form of Eq.~\eqref{eq:series_obs}.

\section{Gradient flow coupling}
\label{sec:gf_nspt}
We employ the gradient flow coupling\cite{Narayanan_2006,L_scher_2010} for the renormalized coupling,
which has emerged as an important tool in both lattice and continuum field theories.
Before discussing the gradient flow coupling in the context of NSPT,
we will first define the gradient flow coupling and explain its relationship with the \MSbar coupling in continuum theory.

\subsection{Gradient flow coupling in the continuum theory}
\label{subsec:gf_coupling_continuum}

The gradient flow coupling, $\lambda_\rho(\mu)$, is defined\cite{Zoltan_Fodor_2012} as
\begin{align}
    \lambda_\rho(\mu)  \equiv& \left. \frac{1}{\mathcal{N}(t)} \expval{ t^2 E(t)} \right|_{\mu^2 t=\rho},
    \label{eq:lgf_continuum}
    \\
    E(t) =& -\frac12 \Tr\qty(G_{\mu\nu}(x,t)G_{\mu\nu}(x,t)),
\end{align}
where $\rho$ denotes an arbitrary parameter; $\mu$ represents the renormalization scale; 
and $E(t)$ denotes the energy density comprised of the field strength tensor $G_{\mu\nu}(x,t)$ evaluated using the flowed gauge potential $B_\mu(x,t)$.
The flowed gauge potential is the solution of the diffusion equation, which is the so-called gradient flow equation\cite{Narayanan_2006,L_scher_2010}:
\begin{align}
    \frac{\partial B_\mu(x,t)}{\partial t} =& D_\nu G_{\nu\mu}[B;t] ,\quad B_\mu(x,0) =A_\mu(x),
    \\
    G_{\mu\nu}(x,t) =& \partial_\mu B_\nu(x,t) - \partial_\nu B_\mu(x,t) + [B_\mu(x,t),B_\nu(x,t)],
\end{align}
where $D_\mu = \partial_\mu + [B_\mu,\cdot]$ denotes the covariant derivative in the adjoint representation.
The extra coordinate $t$, referred to as the flow time, defines the renormalization scale as $\mu=\sqrt{\rho/t}$, as indicated in Eq.~\eqref{eq:lgf_continuum}.
The normalization constant $\mathcal{N}(t)$ is defined ensure that it reproduces the tree-level 
't~Hooft coupling in the perturbation theory and yields $\mathcal{N}(t)=3/(128\pi^2)$ with the continuum perturbation theory in the large-$N$ limit.

The relation between the gradient flow coupling and the \MSbar coupling has been derived 
for the SU($N$) Yang--Mills theory up to the two-loop order in Refs.~\citen{Harlander_2016,Artz_2019}.
This relation for 't~Hooft couplings in the large-$N$ limit becomes
\begin{align}
    \lambda_\rho(\mu) =& \lambda_S(\mu) + \bar{e}_1\lambda_S(\mu)^2 + \bar{e}_2\lambda_S(\mu)^3 + \cdots,
    \label{eq:lgf_msbar}
\end{align}
where $\lambda_S(\mu)$ denotes the \MSbar coupling, with the following coefficients
\begin{align}
    \bar{e}_1 = \frac{e_1}{(4\pi)^2N} = \bar{e}_{1,0} + &\frac12 b_0 L(\rho) ,
    \qquad
    \bar{e}_{1,0} = \frac{1}{16\pi^2}\qty(\frac{52}{9} + \frac{22}{3}\ln2 - 3\ln3 ),
\label{eq:anl_e1}
    \\ 
    &b_0 = \frac{1}{16\pi^2}\frac{2\cdot11}{3},
\label{eq:beta_0}
    \\
    \bar{e}_2 = \frac{e_2}{(4\pi)^4 N^2} = 
    \bar{e}_{2,0} + & \frac12(2b_0\bar{e}_{1,0} + b_1)L(\rho) + \qty(\frac12 b_0 L(\rho) )^2
    ,
\label{eq:anl_e2}
    \\
    \bar{e}_{2,0} = \frac{1}{(16\pi)^2}&27.9786,\qquad
    b_1 = \frac{1}{(16\pi^2)^2}\frac{2\cdot34}{3},
\label{eq:beta_1}
\end{align}
and $L(z)=\ln(2z)+\gamma_E$.
The universal perturbative beta function coefficients, $b_0$ and $b_1$ in Eqs.~\eqref{eq:beta_0} and \eqref{eq:beta_1}, 
satisfy the renormalization group equation\cite{Artz_2019}:
\begin{align}
    \mu \frac{d\lambda_{\rho/S}(\mu)}{d\mu}= -b_0\lambda_{\rho/S}(\mu)^2 -b_1\lambda_{\rho/S}(\mu)^3 + \cdots.
\end{align}

\subsection{Gradient flow with the NSPT on the lattice}
On the lattice, the gradient flow coupling is defined using the flowed link variables $V_\mu(\hat{t})$,
which are smoothed by the gradient flow equation on the lattice\cite{L_scher_2010}:
\begin{align}
    \frac{d V_\mu(\hat{t})}{d\hat{t}}  = - g_0^2\{T^a \partial^a_\mu S[V]\}V_\mu(\hat{t}),
    \qquad
    V_\mu(\hat{t}=0)=U_\mu,
\label{eq:flow_equation_lattice}
\end{align}
where $\hat{t}=t/a^2$ denotes the dimensionless flow time, 
and $\partial^a_\mu$ represents the Lie derivative in the direction of the generator $T^a$\cite{L_scher_2010}.
The explicit form of the gradient flow equation for the TEK model is obtained after substituting Eq.~\eqref{eq:tek_action} into Eq.~\eqref{eq:flow_equation_lattice}:
\begin{align}
    \frac{dV_\mu(\hat{t})}{dt} =& \frac12 F_\mu[V;\hat{t}]V_\mu(\hat{t}),\qquad
    V_\mu(\hat{t}=0) = U_\mu,
    \label{eq:flow_equation_non_perturbative} 
\end{align}
where
\begin{align}
    F_\mu[V;\hat{t}] =& \left(S_\mu - S_\mu^\dag \right) -\frac{1}{N}\mathrm{Tr}\left(S_\mu - S_\mu^\dag \right),\\
    S_\mu =&
    V_\mu \sum_{\nu\neq\mu}\left( z_{\mu\nu}V_\nu  V_\mu^\dag  V_\nu^\dag + z_{\nu\mu}V_\nu^\dag V_\mu^\dag  V_\nu\right).
\end{align}

The gradient flow coupling on the lattice is also defined nonperturbatively\cite{ramos_2013_tgf,ramos_2014_tgf} with the energy density $E(\hat{t})$ as follows
\begin{align}
    \lambda_{\rho,W/C} = \frac{1}{\mathcal{N}_{W/C}(\hat{t})}\expval{\frac{\hat{t}^2 E_{W/C}(\hat{t})}{N}},
\label{eq:def_lgf}
\end{align}
where the subscript $W/C$ corresponds to different definitions of the lattice energy density operator.
Two standard choices are available for the energy density operator on the lattice: the Wilson plaquette-type energy density $E_W$ and clover-type energy density $E_C$:
\begin{align}
    E_W(\hat{t}) =& \Tr\qty(I - z_{\mu\nu}V_\mu(\hat{t}) V_\nu(\hat{t}) V_\mu^\dag(\hat{t}) V_\nu^\dag(\hat{t})),
    \label{eq:energy_density_w}
    \\
    E_C(\hat{t}) =& \frac{1}{2} \sum_{\mu\neq\nu}\Tr\qty( G_{\mu\nu}[V;\hat{t}]G_{\mu\nu}[V;\hat{t}] ),
    \label{eq:energy_density_c}
\end{align}
where
\begin{align}
    G_{\mu\nu}[V;\hat{t}] \equiv& -\frac{i}{8}\qty(C_{\mu\nu}[V;\hat{t}] - C_{\mu\nu}^\dag[V;\hat{t}] ),
    \\
    C_{\mu\nu}[V;\hat{t}] \equiv& z_{\mu\nu}\qty( V_\mu V_\nu V_\mu^\dag V_\nu^\dag + V_\mu^\dag V_\nu^\dag V_\mu V_\nu  + V_\nu^\dag V_\mu V_\nu V_\mu^\dag  + V_\nu V_\mu V_\nu^\dag V_\mu^\dag ).
\end{align}
The normalizing factor $\mathcal{N}_{W/C}(\hat{t})$ is detailed in \ref{sec:tree_level_largen}.

In the NSPT framework, the gradient flow equation is implemented by expanding Eq.~\eqref{eq:flow_equation_lattice} in terms of the coupling constant.
The flowed link variable $V_\mu(\hat{t})$ is expanded as
\begin{align}
    V_\mu(\hat{t}) = \sum_{k=0}^\infty \lambda_0^{k/2} V_\mu^{(k)}(\hat{t}),
\label{eq:series_vmu}
\end{align}
where $V_\mu^{(0)}$ represents the classical vacuum $\Gamma_\mu$.
Substituting Eq.~\eqref{eq:series_vmu} in the gradient flow equation, we obtain the perturbatively expanded gradient flow equation:
\begin{align}
    \frac{d V_\mu^{(k)}(\hat{t})}{dt} =& \frac12 \qty( F_\mu[V;\hat{t}] \star V_\mu(\hat{t}) )^{(k)}
    ,\qquad
    V_\mu^{(k)}(\hat{t}=0) = U_\mu^{(k)},
    \label{eq:flow_equation_nspt} 
\end{align}
where
\begin{align}
    F_\mu^{(k)}[V;\hat{t}] =& \left(S_\mu^{(k)} - S_\mu^{(k)\dag} \right) -\frac{1}{N}\mathrm{Tr}\left(S_\mu^{(k)} - S_\mu^{(k)\dag} \right)
    ,\\
    S_\mu^{(k)} =&
    \left(V_\mu \star \sum_{\nu\neq\mu}\left( z_{\mu\nu}V_\nu \star V_\mu^\dag \star V_\nu^\dag + z_{\nu\mu}V_\nu^\dag \star V_\mu^\dag \star V_\nu\right) \right)^{(k)},
\end{align}
for $ k=1,\cdots,\infty$.
The lowest order coefficient $V_\mu^{(0)}(\hat{t})$ does not evolve and remains fixed as $V_\mu^{(0)}(\hat{t})=U_\mu^{(0)}=\Gamma_\mu$.
The $\star$-product represents the convolution product of matrix polynomials, defined as\cite{Gonz_lez_Arroyo_2022}
\begin{align}
    C^{(k)} = \qty(A \star B)^{(k)} \equiv&  \sum_{l=0}^{k} A^{(k-l)} B^{(l)},
\end{align}
for matrix polynomials 
\begin{align}
A = \sum_{k=0}^\infty \lambda_0^{k/2} A^{(k)},\quad
B = \sum_{k=0}^\infty \lambda_0^{k/2} B^{(k)},\quad
C = \sum_{k=0}^\infty \lambda_0^{k/2} C^{(k)}.
\end{align}
The energy density in the NSPT framework can be derived similarly by substituting Eq.~\eqref{eq:series_vmu} into Eqs.~\eqref{eq:energy_density_w} and \eqref{eq:energy_density_c}, where the matrix products are replaced by the $\star$-product.

The $k$-th order coefficient $E_{W/C}^{(k)}(\hat{t})$ of the energy density in Eq.~\eqref{eq:def_lgf} is estimated as the stochastic average:
\begin{align}
    \expval{E_{W/C}^{(k)}(\hat{t})[V_\mu^{(0)},\cdots,V_\mu^{(k)}]} \simeq
    \frac{1}{N_\mathrm{sample}} \sum_{i=1}^{N_\mathrm{sample}}E_{W/C}^{(k)}(\hat{t})[V_{\mu,i}^{(0)},\cdots,V_{\mu,i}^{(k)}],
\end{align}
where $V_{\mu,i}^{(k)}(\hat{t})$ denote the coefficients of the flowed link variables evolved from the initial values $(U_{\mu,i}^{(0)},\cdots,U_{\mu,i}^{(k)})$ for the $i$-th sample.
The flow equation Eq.~\eqref{eq:flow_equation_nspt} is numerically integrated using an integration scheme.
This approach allows us to compute the perturbative expansion of the gradient flow coupling from Eq.~\eqref{eq:def_lgf}.

\subsection{Gradient flow coupling in terms of the lattice bare coupling}
\label{subsec:gf_lattice}
In this subsection we derive the gradient flow coupling in terms of the lattice bare coupling constant $\lambda_0$.
This expression is obtained by combining existing results Eq.~\eqref{eq:lgf_msbar} with the relation between the \MSbar and lattice bare coupling\cite{L_scher_1995} in the large-$N$ limit.
The \MSbar coupling can be expressed in terms of the lattice bare coupling as follows
\begin{align}
    \lambda_S(\mu) =& \lambda_0 + c_1(\mu a){\lambda_0}^2 + c_2(\mu a) {\lambda_0}^3 + \cdots,
    \label{eq:msbar_bare}
\end{align}
where the coefficients are
\begin{align}
    &c_1(\mu a) =  -b_0\ln(\mu a) + \frac{k_1}{4\pi},
    \quad 
    k_1=2.135730074078457(2),
\label{eq:anl_c1}
    \\
    &c_2(\mu a) =  c_1^2(\mu a) -b_1\ln(\mu a) + \frac{k_3}{16\pi^2},
    \quad 
    k_3=1.24911585(3),
\label{eq:anl_c2}
\end{align}
in the large-$N$ limit and $\lambda_0$ denotes the bare lattice 't~Hooft coupling of the Wilson gauge action.
The universal perturbative beta function coefficients, $b_0$ and $b_1$, are specified in Eqs.~\eqref{eq:beta_0} and \eqref{eq:beta_1}.
By substituting Eq.~\eqref{eq:msbar_bare} into Eq.~\eqref{eq:lgf_msbar},
we can express the gradient flow coupling in terms of the lattice bare coupling as
\begin{align}
    \lambda_\rho(\mu) = \lambda_0 + r_1(\hat{t})\lambda_0^2 + r_2(\hat{t})\lambda_0^3 + \cdots
    \label{eq:lgf_bare}
\end{align}
\begin{align}
    r_1(\hat{t}) & =  b_0\qty(\log(\sqrt{2\hat{t}} )+ \frac{\gamma_E}{2}) + f_1,
    \label{eq:anl_r1}  \\
    r_2(\hat{t}) & =  r_1(\hat{t})^2 + b_1\qty(\log(\sqrt{2\hat{t}}) + \frac{\gamma_E}{2}) + f_2,
    \label{eq:anl_r2} 
\end{align}
\begin{align}
    f_1 &= 0.21786204948590122,\qquad  f_2 = 0.006737113, 
\end{align}
where $\mu = \sqrt{\rho/t}$.
We anticipate that in the large-$N$ limit, the lattice bare coupling of the SU($N$) Wilson gauge action
and that of the SU($N$) TEK model will be identical.

This paper details the NSPT calculation for the gradient flow coupling for
the TEK model in the large-$N$ limit, incorporating the three-loop coefficient $r_3(\hat{t})$
 which is not explicitly detailed in Eq.~\eqref{eq:lgf_bare}.
The coefficients $r_1(\hat{t})$ and $r_2(\hat{t})$ are detailed in the following section.
Specifically, the one-loop coefficient $r_1(\hat{t})$ is reproduced using NSPT within our current setup;
however, owing to large statistical errors, precise determination of the two- and three-loop perturbative beta coefficients,
$r_2(\hat{t})$ and $r_3(\hat{t})$, is challenging.
We will detail the sample size for the two-loop coefficient with sufficient precision based on a detailed study of the variance of $r_2(\hat{t})$.

\begin{table}[t]
    \tbl{Parameters for the NSPT algorithm and the TEK model.}{
    \begin{tabular}{ccccccccc}
    \toprule
     $\hat{L}$ & $N$ & $k$ & $|\bar{k}|$ & $\theta=|\bar{k}|/L$ & $\tau$ & $N_{\mathrm{MD}}$ & Statics & $T_\mathrm{GF}$ [sec/conf]    
     \\ \colrule
     17 & 289 & 5 & 7 & 0.41176  & 1.0 & 32 & 5931 & 232 
     \\
     21 & 441 & 13 & 8 & 0.38099 & 1.0 & 32 & 3790 & 620
     \\
     23 & 529 & 7 & 10 & 0.43479 & 1.0 & 32 & 2600 & 1091
     \\ 
     \botrule
     \end{tabular}
    \label{tab:nspt_parameter}}
\end{table}

\section{Numerical results}
\label{sec:numerical_result}
We accumulated the perturbative configuration using the NSPT algorithm detailed in Ref.~\citen{Gonz_lez_Arroyo_2019} and 
evaluated the perturbative coefficients of the gradient flow coupling by following the procedure detailed in the previous section.
The parameters of the NSPT algorithm and the TEK model are outlined in Table~\ref{tab:nspt_parameter}. 
Specifically, we set the partial momentum refreshment parameter to $c_1 = \exp(-\gamma\Delta\tau)$ with $\gamma=2.0$ 
and the leap flog step size as $\Delta\tau=1/32$.
Additional details regarding these parameters can be found in Ref.~\citen{Gonz_lez_Arroyo_2019}.
To take the large-$N$ limit we used three matrix sizes, $N=289,441,529$,
for the SU($N$) TEK model with fixing the phase parameter at $\theta\simeq0.40$ to ensure a smooth large-$N$ limit.
The number of configurations generated for each setup is also shown in Table~\ref{tab:nspt_parameter}, with each sample separated by $\tau=4$.

\renewcommand{\figscale}{0.38}
\begin{figure}[t]
    \centering
    \includegraphics[clip,scale=\figscale]{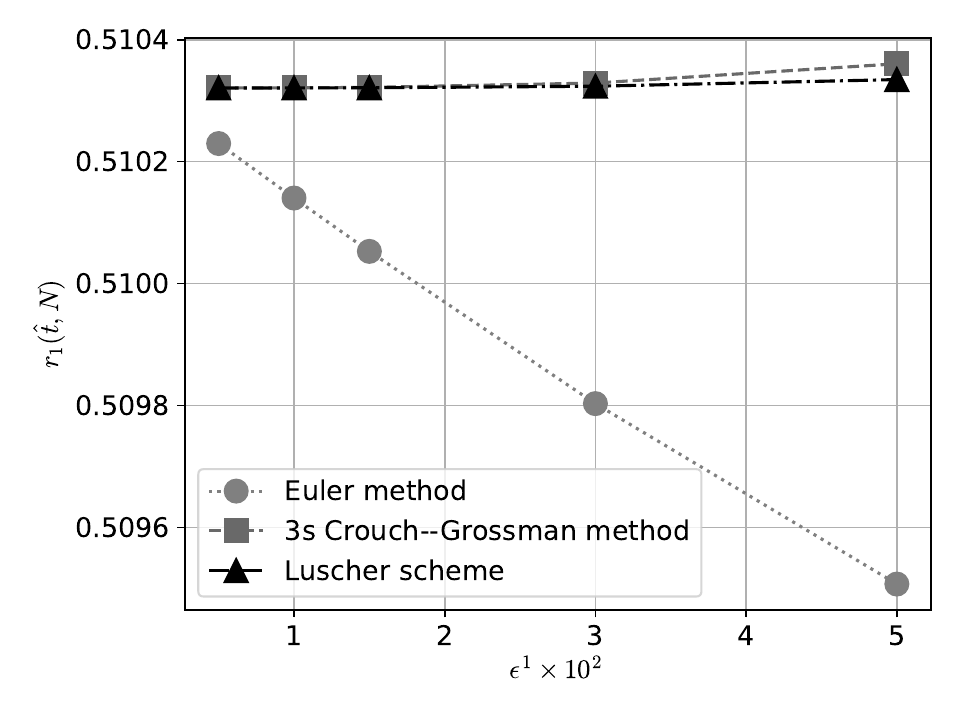}   \hfill
    \includegraphics[clip,scale=\figscale]{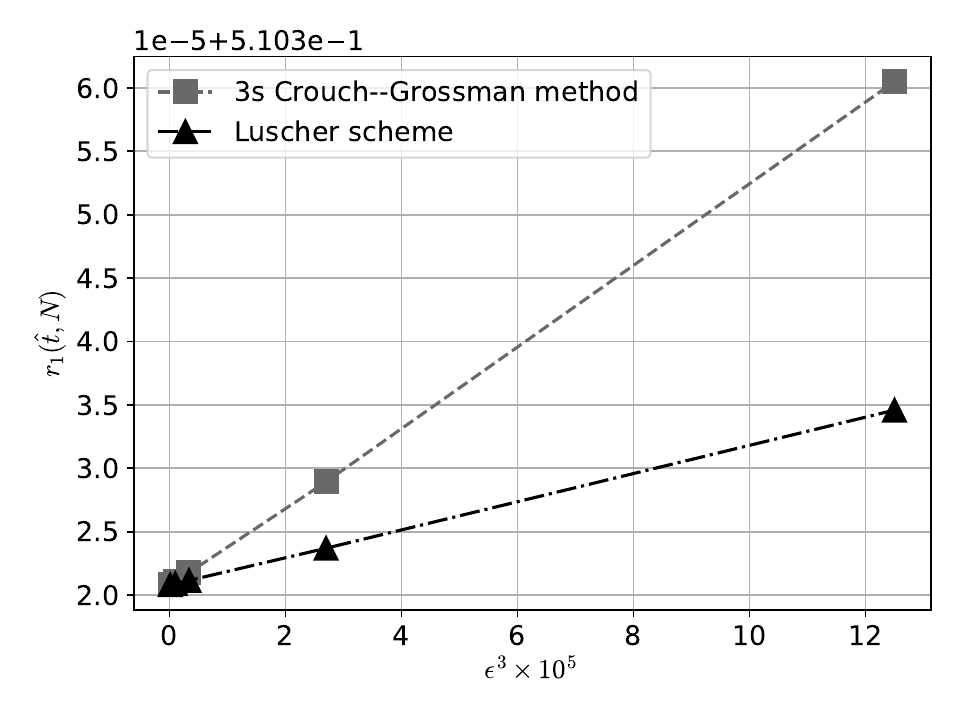}
    \includegraphics[clip,scale=\figscale]{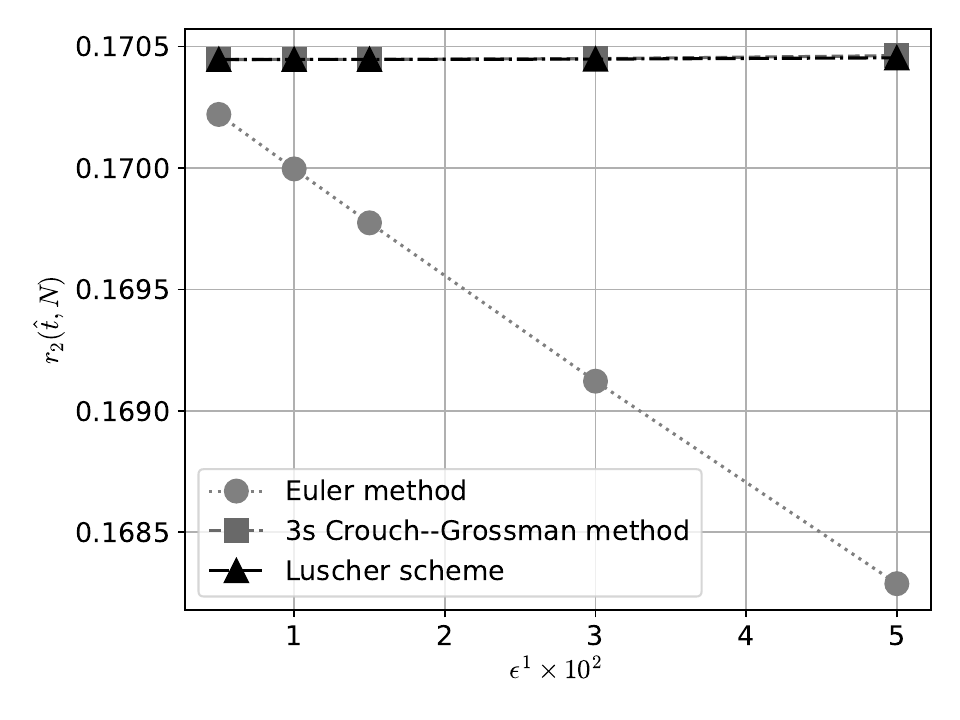}   \hfill
    \includegraphics[clip,scale=\figscale]{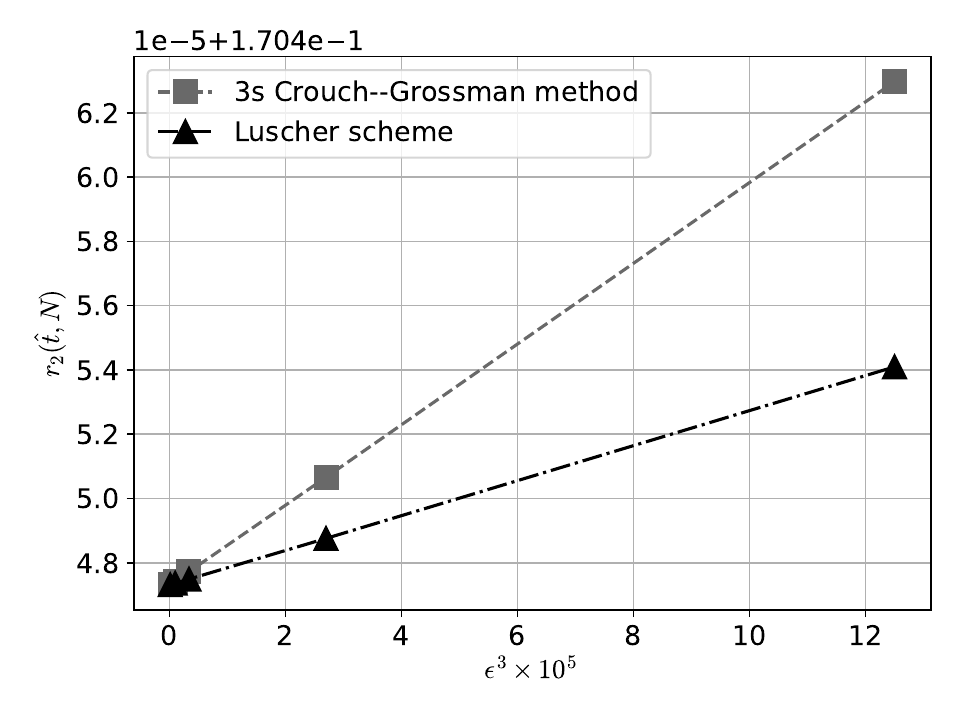}
    \includegraphics[clip,scale=\figscale]{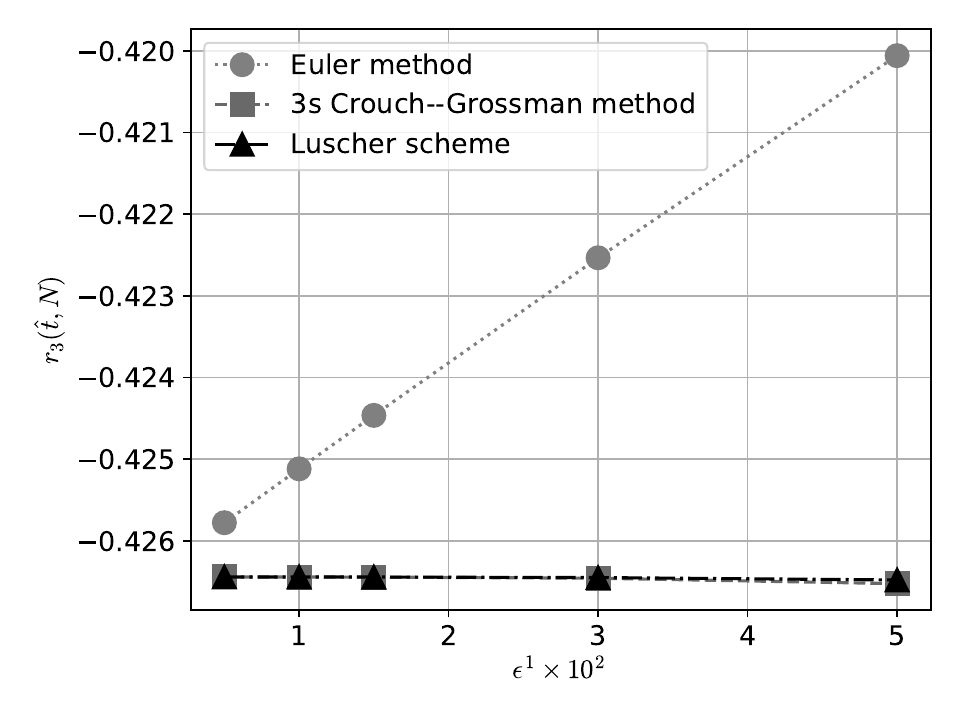}   \hfill
    \includegraphics[clip,scale=\figscale]{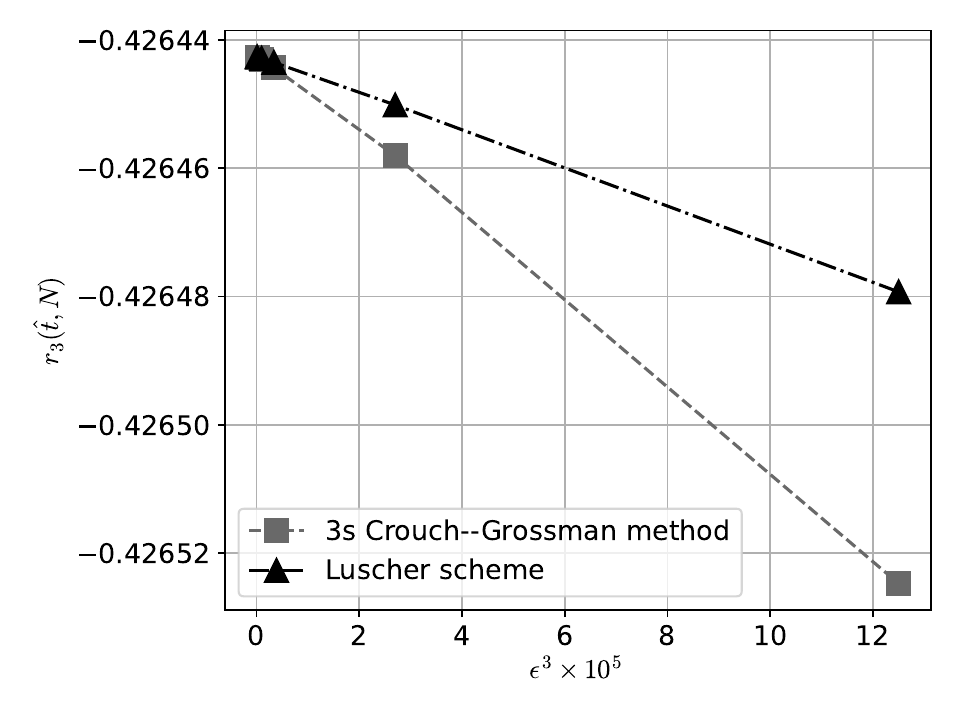}
    \caption{Scaling test of the coefficients using the Euler method, three-stage Crouch--Grossman method, 
             and L\"uscher scheme at $\hat{t}=9.0$ on a single configuration with $k=7,N=529$ for the TEK model.
             The column presents the coefficients as a function of $\epsilon$, while
             the right column shows the results, excluding those for the Euler method, as a function of $\epsilon^3$.}
    \label{fig:test_scale}
\end{figure}

To integrate the gradient flow equation Eq.~\eqref{eq:flow_equation_nspt}, we employed
a numerical integration scheme with a finite discrete step size.
Among the several numerical integration schemes developed to date, we tested the following three methods to determine
the appropriate step size where the integration error becomes smaller than the statistical error:
the Euler method, the L\"uscher scheme\cite{L_scher_2010}, and a three-stage Crouch--Grossman method.
Figure~\ref{fig:test_scale} shows the dependence of the perturbative coefficients of 
the gradient flow coupling on the step size at a flow time $\hat{t}=9.0$, using a single configuration with $k=7, N=529$.
As anticipated, the L\"uscher scheme (up-triangles) and the three-stage Crouch--Grossman method (boxes)
are accurate up to errors of order $\order{\epsilon^3}$ and exhibit faster convergence to the zero step size limit
compared to the Euler method (circles).
Similar results were observed for smaller values of $N$, 
which led us to adopt the L\"uscher scheme with $\epsilon=0.01$ for further calculations 
owing to its computational efficiency and lower integration errors compared to the other schemes.

Numerical computation programs for generating configurations and computing the gradient flow coupling
were developed using Fortran2003/2008 combined with CUDA\cite{nvidia_cuda} to leverage graphics processing unit accelerators,
which offer a cutting-edge performance for matrix-matrix multiplication in the convolution product of NSPT link variables.
We employed two high-performance computing facilities;
Cygnus at the CCS University of Tsukuba\cite{tsukuba_cygnus} and 
ITO System-B at Kyushu University\cite{kyushu_ito}.
We measured the computation time $T_\mathrm{GF}$ required to numerically integrate the flow time for 
$N_\mathrm{step}=1500$ (equivalently to $\hat{t}$=15) on one configuration, as detailed in Table~\ref{tab:nspt_parameter}. 
The time was measured on the Cygnus system.
The time required for configuration generation is approximately five times lower 
than that required for integrating the gradient flow equation with our selected step sizes.
Hence, the total computation time, including the time for configuration generation,
is dominated by $T_{GF}$ the time for numerically integrating the gradient flow equation.

\subsection{Gradient flow coupling}
\label{subsec:result_lgf}
We computed the gradient flow coupling as follows:
\begin{align}
    \lambda_\rho(\mu) = \lambda_0 + r_1(\hat{t},N)\lambda_0^2 + r_2(\hat{t},N)\lambda_0^3 + r_3(\hat{t},N)\lambda_0^4,
    \label{eq:lgf_finiteN}
\end{align}
up to the three-loop level at each finite $N$ values $289$, $441$, and $529$. 
The gradient flow coupling was evaluated at every flow time separated by $\Delta\hat{t}=15\epsilon=0.15$.
We extrapolated the finite-$N$ results $r_i(\hat{t},N),\ (i=1,2,3)$ to the large-$N$ limit as a function
of $r_i(\hat{t},N)=r_i(\hat{t}) + s_{i}(\hat{t})/N^2$ at fixed flow times $\hat{t}$.
The results at various flow times are summarized in \ref{sec:some_result}.
In particular, the results revealed that the $1/N^2$ extrapolation is consistent with reasonable chi-square values.
Figure~\ref{fig:r1_largeN_wide} illustrates the perturbative coefficients $r_1(\hat{t})$ and $r_2(\hat{t})$ in the large-$N$ limit.
Notably, we only plot the results using the clover-type operator because the results with the Wilson-type operator are statistically identical to those derived using the clover-type operator in this flow time region.
The left and right panels of Fig.~\ref{fig:r1_largeN_wide} present $r_1(\hat{t})$ and $r_2(\hat{t})$, respectively.
The dashed lines denote the analytic curves from Eqs.~\eqref{eq:anl_r1} and \eqref{eq:anl_r2},
while the crosses with bars, where statistical errors are evaluated using the bootstrap method, correspond to the NSPT results.
We observe that the NSPT results are consistent with the analytic predictions in the mid-flow time region
$\hat{t}\in[2.1,6.3]$ for $r_1(\hat{t})$ and in the large flow time $\hat{t}>3.0$ for $r_2(\hat{t})$.
In the small flow time region, the finite cut-off error of $\order{a^2/t}$ is anticipated because the TEK model and observable possess no $\order{a}$ errors.
In contrast, the finite-$N$ error of order $\order{\hat{t}^2/N^2}$, corresponding to the finite volume effect 
of order $\order{t^2/V}$ with the effective volume $V=(a\sqrt{N})^4$ for the TEK model, is anticipated in the large flow time region. 
Details regarding these systematic errors, derived using the tree-level analysis, are presented in \ref{sec:tree_level_largen}.

\renewcommand{\figscale}{0.39}
\begin{figure}[t]  
\centering
    \includegraphics[clip,scale=\figscale]{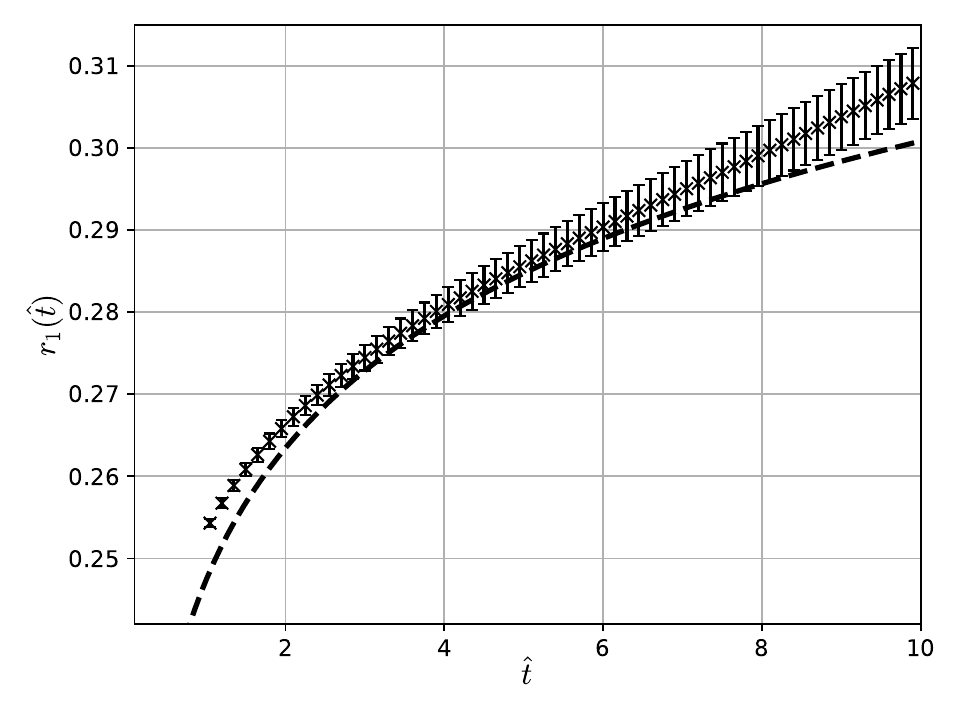}\hfill
    \includegraphics[clip,scale=\figscale]{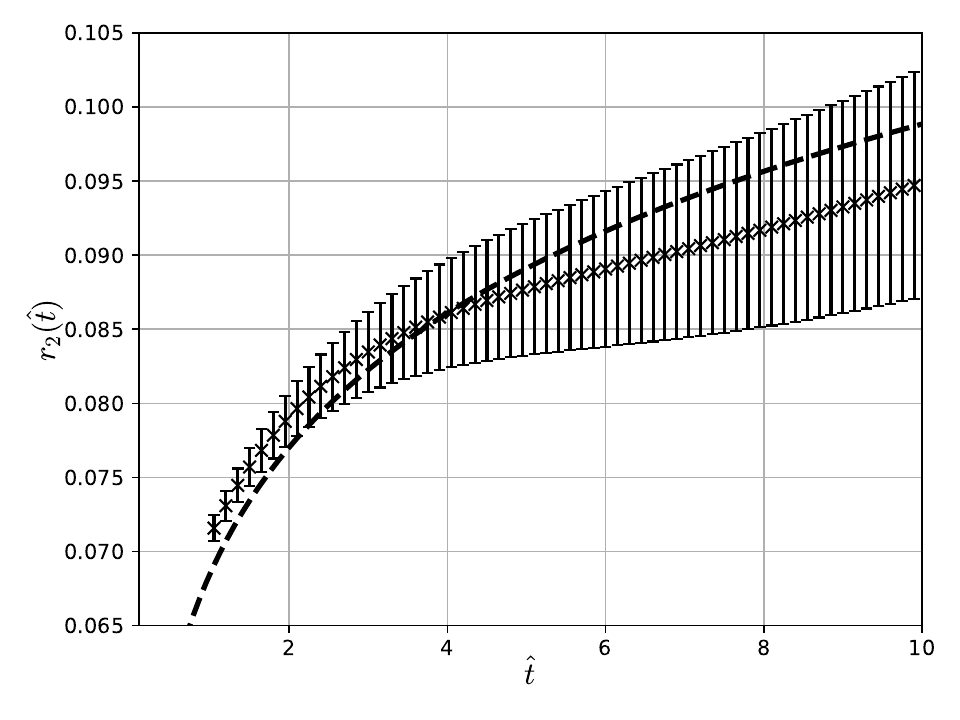}
    \caption{
    Flow time dependence of coefficients $r_1(\hat{t})$ and $r_2(\hat{t})$ in the large-$N$ limit.
    Crosses and dashed lines represent extrapolation results and continuum analytic expressions, respectively.
    }
    \label{fig:r1_largeN_wide}
\end{figure}

For the three-loop coefficient $r_3(\hat{t})$, large statistical error prevents us from extracting its precise flow time dependence within the current statistical sample size.
Thus, we do not present the results for $r_3(\hat{t})$.
Instead, we specify the sample size required to extract the three-loop beta function from the flow time dependence in Section~\ref{sec:summary_and_outlook}.
In the next subsection we analyze the flow time dependence of $r_1(\hat{t})$ and $r_2(\hat{t})$ in more detail to assess their consistency with the analytic results.

\subsection{Analysis of flow time dependence}
\label{subsec:analysis_flow}
We fitted the results of the coefficient $r_1(\hat{t})$ in the large-$N$ limit using the two fitting functions:
\begin{align}
    f(\hat{t}) =& B_0\qty(\log(\sqrt{2\hat{t}}) + \frac{\gamma_E}{2}) + F_1,
    \label{eq:fit_r1_fx}
    \\
    g(\hat{t}) =& B_0\qty(\log(\sqrt{2\hat{t}}) + \frac{\gamma_E}{2}) + F_1 + \frac{A_0}{\hat{t}},
    \label{eq:fit_r1_gx}
\end{align}
where $B_0$, $F_1$, and $A_0$ denote fitting parameters.
Both functions are based on Eq.~\eqref{eq:anl_r1};
however, Eq.~\eqref{eq:fit_r1_gx} includes an additional lattice artifact term, $A_0/\hat{t}$.
The effect of the lattice artifact $\order{a^2/t}$ becomes significant in the small flow time region.
We fitted the flow time dependence of $r_1(\hat{t})$ measured at discrete flow time $\Delta\hat{t}=0.15$ within a specific flow time region.
Given that the coefficients at different flow times are strongly statistically correlated,
we used a least squares fitting method that incorporates the covariance matrix of the correlated data.
To evaluate the covariance matrix, we employed the bootstrap method, with diagonal elements correspoinding to the square of the statistical error.
However, we observe the near zero eigenvalues of the covariance matrix in double precision, which caused unrealistic fitting when using the least squares approah.
To address this issue, we applied the cut-off method\cite{yoon_2012}, where small eigenvalues below a cut-off threshold were excluded from the covariance matrix.

\begin{figure}[t]  
    \centering
        \includegraphics[clip,scale=\figscale]{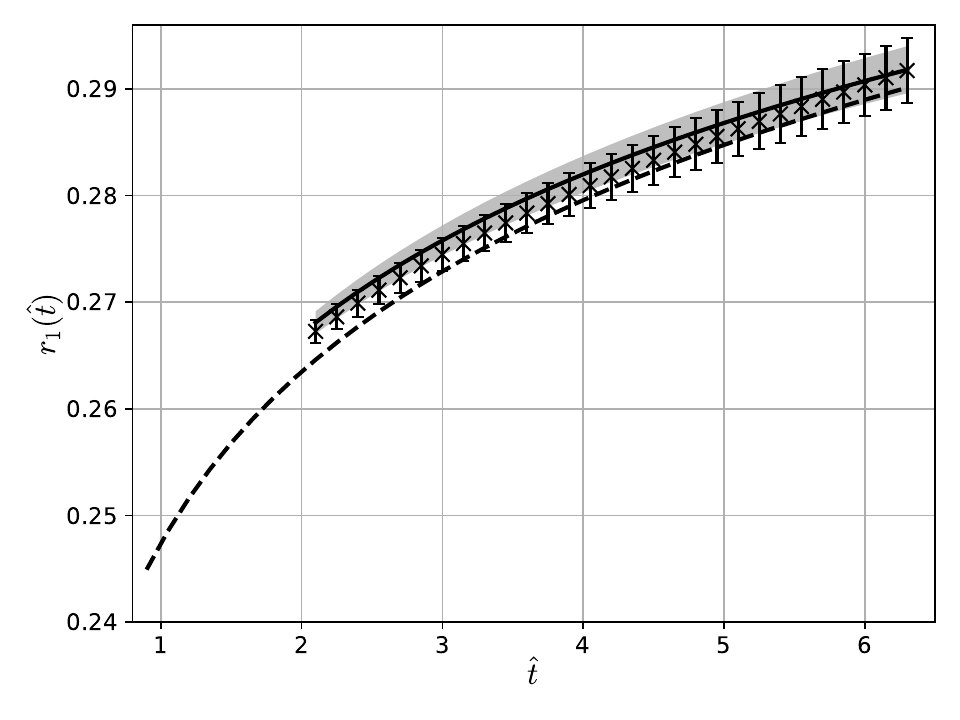}\hfill
        \includegraphics[clip,scale=\figscale]{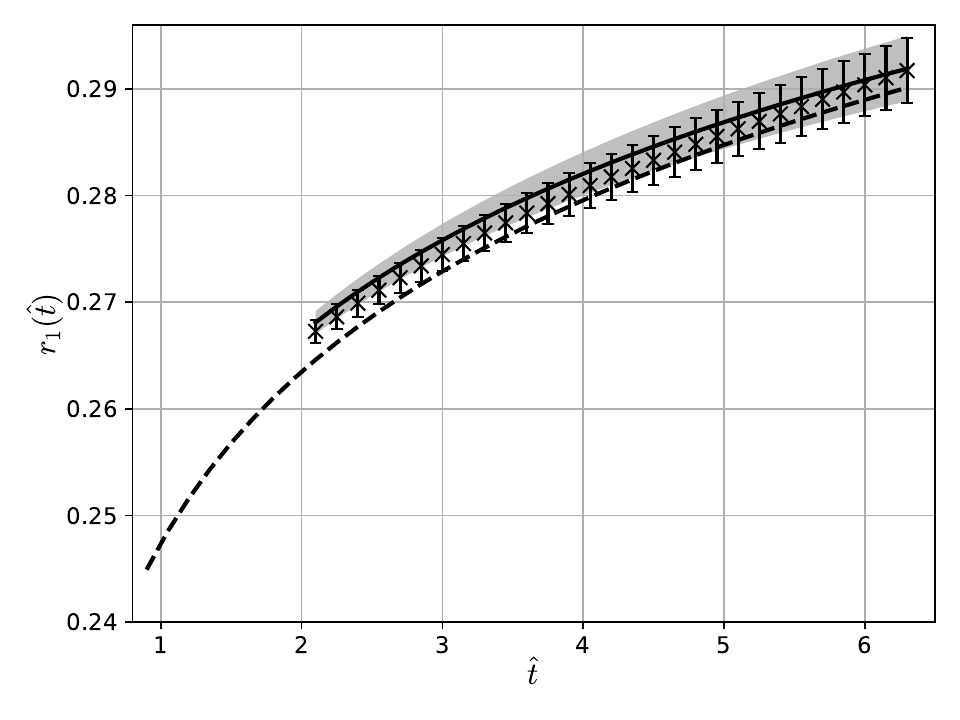}
        \includegraphics[clip,scale=\figscale]{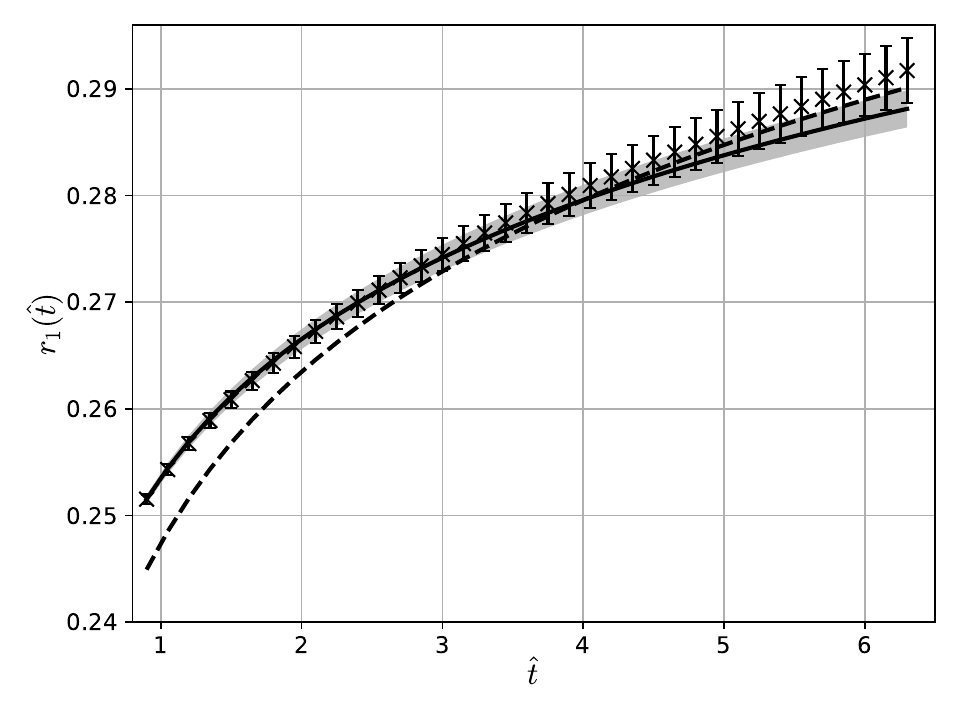}\hfill
        \includegraphics[clip,scale=\figscale]{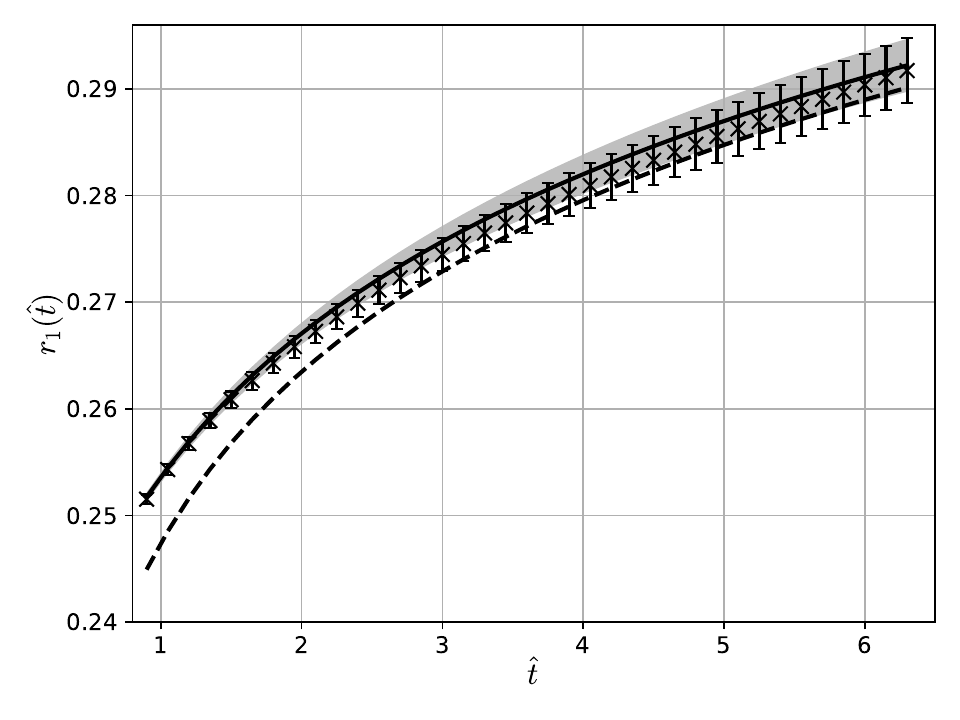}
        \caption{Flow time dependence of coefficient $r_1(\hat{t})$ in the large-$N$.
        Top (bottom) panels present the fitting results in the flow time region $\hat{t}\in[2.1,6.3]$ ($\hat{t}\in[0.9,6.3]$).
        The definitions of crosses and dashed lines are identical to those in Fig.~\ref{fig:r1_largeN_wide}.
        Solid lines represent the fitting results with $f(x)$ (left panels) and $g(x)$ (right panels).
        }
        \label{fig:r1_largeN}
    \end{figure}

Figure~\ref{fig:r1_largeN} illustrates the fitting results for the one-loop coefficient $r_1(\hat{t})$.
In this figure, the definitions of crosses and dashed lines are identical to those in Fig.~\ref{fig:r1_largeN_wide}.
The solid lines with shaded region represent the fitting results obtained using the cut-off method.
We employ two fitting regions, namely $\hat{t}\in[2.1,6.3]$ (top panels) and $[0.9,6.3]$ (bottom panels), to examine the effect of the lattice artifact $\order{a^2/t}$ term.
Notably, the finite volume effect $\order{t^2/V}$ intensifies with flow time.
Based on the tree-level analysis detailed in \ref{sec:tree_level_largen}, we determined that $N>787$ is essential to suppress this correction to under $5\%$ at $\hat{t}=7.0$.
Given that we only have three datasets at $N=289$, $441$, and $529$ which do not satisfy this condition, our simple linear extrapolation in $1/N^2$ is not applicable in the large flow time region.
Hence, we discard the results for the large flow time $\hat{t}>6.3$.
In the small time region $\hat{t}<0.9$, the effects of lattice artifacts $\order{a^2/t}$ and $\order{a^4/t^2}$ are dominant;
hence, we use the results for $\hat{t}\geq0.9$.

\begin{table}[t]
    \tbl{Fitting parameters of coefficient $r_1(\hat{t})$ in the flow time region $\hat{t}\in[2.1,6.3]$.
    $\kappa$ denotes the cut-off size for the eigenvalue of the covariance matrix.
    The results derived using the cut-off method are plotted in Fig.~\ref{fig:r1_largeN}.}{
    \begin{tabular}{ccllll} \toprule
        Fit function & $\kappa$ & $B_0$ & $F_1$ & $A_0$ & $\chi^2/N_\mathrm{dof}$\\ \colrule
        \multirow{1}{*}{$f(x)$} 
              & $10^{-10}$ & 0.04315(222) & 0.22466(139)  & --           & 12.3 \\
        \multirow{1}{*}{$g(x)$} 
              & $10^{-10}$ & 0.04349(748) & 0.22419(1009) & 0.00030(634) & 12.1 \\ \colrule
        Analytical value & & 0.046439     & 0.217862 & -- & -- \\
    \botrule
    \end{tabular}
    \label{tab:r1_fit_midle}}
\end{table}

\begin{table}[t]
    \tbl{Same as Table.~\ref{tab:r1_fit_midle}, but the fit region is $\hat{t}\in[0.9,6.3]$.}{
    \begin{tabular}{ccllll} \toprule
        Fit function & $\kappa$ & $B_0$ & $F_1$ & $A_0$ & $\chi^2/N_\mathrm{dof}$\\
        \colrule
        \multirow{1}{*}{$f(x)$}
         & $10^{-8}$ & 0.03762(144) & 0.22959(60) & -- & 3.2 \\
        \multirow{1}{*}{$g(x)$}
         & $10^{-9}$ & 0.04725(349) & 0.21778(337) & 0.00577(139) & 4.2 \\ \colrule
        Analytical value & & 0.046439 & 0.217862 & -- & -- \\
    \botrule
    \end{tabular}
    \label{tab:r1_fit_small}}
\end{table}

The fitting results for $B_0$, $F_1$, and $A_0$ are detailed in Tables~\ref{tab:r1_fit_midle} and \ref{tab:r1_fit_small}.
We observe larger values of $\chi^2/N_\mathrm{dof}$ for the fitting results in the region $\hat{t}\in[2.1,6.3]$ (Table~\ref{tab:r1_fit_midle})
compared to those in the region $\hat{t}\in[0.9,6.3]$ (Table~\ref{tab:r1_fit_small}).
The coefficient $A_0$ for the lattice artifact has no effect in the region $\hat{t}\in[2.1,6.3]$;
however, it exhibits non-zero contributions in the region $\hat{t}\in[0.9,6.3]$.
As indicated in Table~\ref{tab:r1_fit_small}, the fitting results with the fit function $g(x)$ are consistent with the analytic values of $B_0$ and $F_1$,
indicating the existence of a flow time window where the lattice artifact is sufficiently under control.

\begin{figure}[t]  
    \centering
        \includegraphics[clip,scale=\figscale]{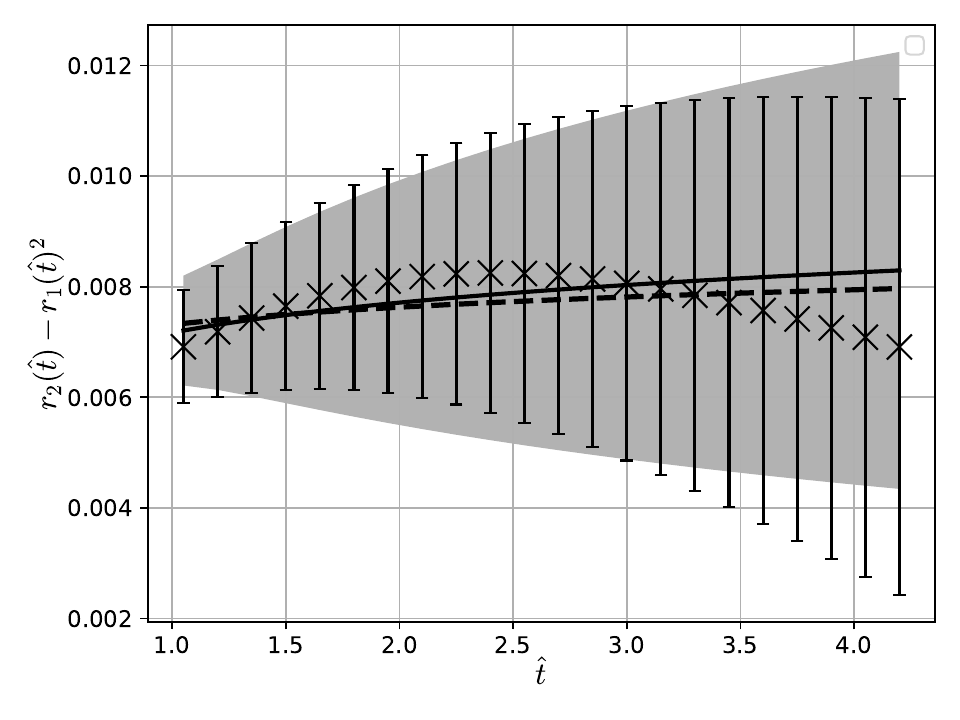}
        \caption{Same as Fig.~\ref{fig:r1_largeN}, but for $r_2(\hat{t})-r_1(\hat{t})^2$.}
        \label{fig:r2_largeN}
\end{figure}

Figure~\ref{fig:r2_largeN} presents the two-loop results for the subtracted coefficient $r_2(\hat{t})-r_1(\hat{t})^2$.
In this figure, the crosses denote the two-loop coefficient in the large-$N$ limit.
To obtain the large-$N$ limit, we first evaluate the subtracted result $r_2(\hat{t},N)-r_1(\hat{t},N)^2$
at finite-$N$ and subsequently extrapolate to the large-$N$ limit.
The dashed line in Fig.~\ref{fig:r2_largeN} represents the analytic coefficient Eq.~\eqref{eq:anl_r2}, while the solid line corresponds to the fitting result obtained using 
\begin{align}
    f(\hat{t}) = B_1\qty(\log(\sqrt{2\hat{t}}) + \frac{\gamma_E}{2}) + F_2,
    \label{eq:fit_r2}
\end{align}
where $B_1$ and $F_2$ denote fitting parameters.
As illustrated in Fig.~\ref{fig:r2_largeN} the values are consistent with the continuum results;
however, the statistical error increases as the subtracted coefficient becomes smaller.
We also observed that obtaining a correlated fit for the data is difficult.
Despite these challenges, we applied an uncorrelated fitting to the data in region $\hat{t}\in[0.9,4.2]$, and the results are summarized in Table~\ref{tab:r2_fit_largeN}.
The large error prevent us from precisely extracting the beta function coefficient.
To accurately extract the beta function coefficient, we need to further reduce the statistical error in the large-$N$ limit.

One approach to reduce the statistical error is to increase the number of samples at each $N=289$, $441$, and $529$.
Furthermore, including several data points for larger $N$ values could help stabilize the flow time fitting, 
as it would provide a broader flow time region where the large-$N$ limit can be safely taken with the controlled $\order{\hat{t}^2/N^2}$ errors.
Moreover, the large-$N$ factorization property could reduce the variance of the coefficients,
yielding smaller statistical errors at larger $N$.

\begin{table}[t]
    \tbl{Fitting parameters of the coefficient $r_2(\hat{t})-r_1(\hat{t})^2$ in the flow time region $\hat{t}\in[2.1,6.3]$.
    An uncorrelated fit was used in this case.
    }{
     \begin{tabular}{clll}\toprule
        Fit function  &    $B_1$       & $F_2$        & $\chi^2/N_\mathrm{dof}$\\ \colrule
            $f(x)$    &   0.00157(491) & 0.00617(293) & 0.032                  \\ \colrule
     Analytical value &   0.00090897   & 0.00673711   & -- \\  
    \botrule
    \end{tabular}
    \label{tab:r2_fit_largeN}}
\end{table}

Subsection~\ref{subsec:factorization} details the estimation of the sample size required for larger $N$ using the current dataset.
The sample size and statistical error are related through the variance of the observable.
In the context of the SU($N$) Yang--Mills theory, the large-$N$ factorization property indicates that the expectation value for an operator product behaves as
\begin{align}
    \expval{O_1 \cdots O_n}
    =\expval{O_1} \cdots \expval{O_n} + \order{\frac{1}{N^2}},
\end{align}
in the large-$N$ limit, where $O_1,\dots,O_n$ denote the gauge invariant operators at different space-time points\cite{Witten_1979}.
Given that the matrix model is based on the large-$N$ factorization property, the same factorization behavior is observed in the current matrix model.
Thus we expect
\begin{align}
    \mathrm{Var}(O) \equiv \expval{O^2} - \expval{O}^2
    =\order{\frac{1}{N^2}}
    \to 0.
\end{align}
The next subsection explicitly assesses the variance of the coefficients of $r_1(\hat{t})$ and $r_2(\hat{t})$ 
at several positive flow times and estimates the sample size based on this variance.

\subsection{Large-$N$ factorization and estimation of sample size}
\label{subsec:factorization}
The large-$N$ factorization property is crucial in the context of the SU($N$) gauge theory in the large-$N$ limit.
This property implies that the variance of an observable becomes zero as we approach the large-$N$ limit.
Previous studies\cite{Gonz_lez_Arroyo_2019,Gonz_lez_Arroyo_2022} have demonstrated the large-$N$ factorization 
of the expectation value of the perturbation coefficient in the matrix model.
However, for flowed operators at a finite flow time, large-$N$ factorization is not trivial.
In this subsection, we analyze the variance of the coefficients at finite $N$ and estimate the required sample size for larger $N$.

Figure~\ref{fig:variance_t6} illustrates the variance at finite flow time $\hat{t}=6.0$.
The left and right panels depict the variances of the coefficients $r_1(\hat{t})$ and $r_2(\hat{t})$, respectively.
The diamonds in the figures correspond to the numerical results at finite $N$ values of $289$, $441$, and $529$.
The solid lines correspond to the results of the simple linear fitting performed using $f(N)=A_0+A_1/N^2$,
while the dashed lines represent the fitting results for $N=529$ obtained using $f(N)= A_1/N^2$ under 
the constraint that the variance vanishes in the large-$N$ limit.
We observe that the variance (down-triangle) at the large-$N$ limit extrapolated using the data for
$N=289$, $441$, and $529$ is finite, which appears to contradict the large-$N$ factorization property.
However, this is not a contradiction because the finite volume correction of $\order{\hat{t}^2/N^4}$ 
can be large at $\hat{t}=6.0$ for smaller $N$ values of $289$, $441$, and $529$.
This point is clarified in \ref{sec:var_ene} based on a tree-level analysis of the variance of the energy density operator.
Thus the simple linear extrapolation fails to obtain the zero variance at large-$N$ from our finite $N$ results in the large flow time region.
We also note that for $\hat{t} \simeq 0$, the variance at the large-$N$ limit is consistent with zero when using the linear extrapolation approach. 
To estimate the required sample size at $N=800$ and flow time $\hat{t}=6.0$ from the variance,
we interpolate the variance using both fitting approaches as indicated by stars in the figures.
These estimates provide the upper and lower bounds for the required number of samples at $N=800$.

\renewcommand{\figscale}{0.38}
\begin{figure}[ht]  
    \centering
    \includegraphics[clip,scale=\figscale]{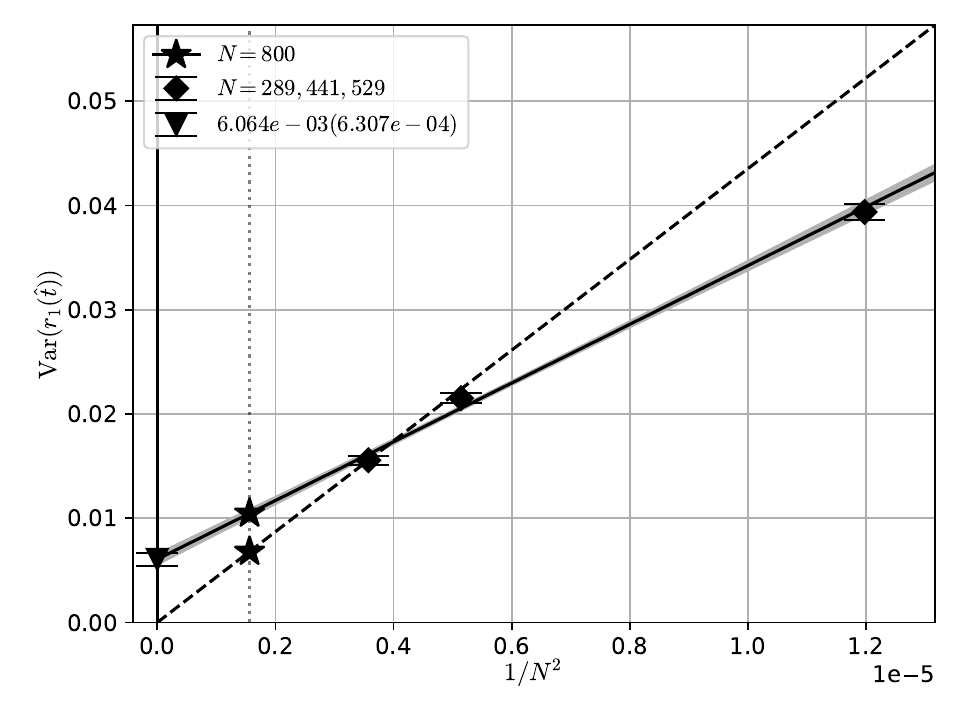}  \hfill
    \includegraphics[clip,scale=\figscale]{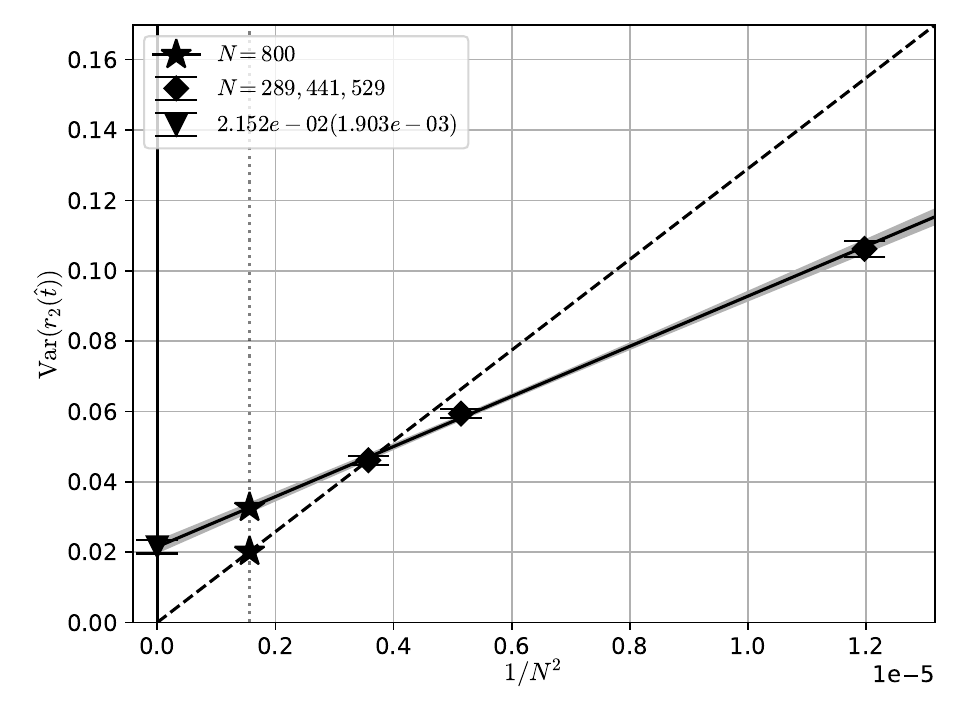}
    \caption{
    Variances of the coefficients $r_1(\hat{t})$ (right) and $r_2(\hat{t})$ (left) at $\hat{t}=6.0$.
    The diamonds denote the numerical results of the variances,
    while down-triangles indicate the results for large-$N$ obtained using simple linear fitting (solid line).
    }
    \label{fig:variance_t6}
\end{figure}

The statistical error $\delta r_i(\hat{t},N)$ is related to the variance $\mathrm{Var}(r_i(\hat{t},N))$ 
and the number of sample $N_\mathrm{sample}$ as follows: 
\begin{align}
    \qty(\delta r_i(\hat{t},N))^2 = \frac{\mathrm{Var}(r_i(\hat{t},N))}{N_\mathrm{sample}}.
\label{eq:statistical_error}
\end{align}
From Eq.~\eqref{eq:statistical_error}, we derive the relationship between the number of samples
and the relative statistical error $R={\delta r_i(\hat{t},N)}/{\expval{r_i(\hat{t},N)}}$ as follows:
\begin{align}
    N_\mathrm{sample} 
    \equiv \frac{\mathrm{Var}(r_i(\hat{t},N))}{\expval{r_i(\hat{t},N)}^2} \frac{1}{R^2}.
\label{eq:estm_nsmp}
\end{align}
To estimate the required sample size for a fixed relative statistical error using Eq.~\eqref{eq:estm_nsmp},
we use the analytic coefficients $\expval{r_i(\hat{t})}$ given in Eqs.~\eqref{eq:anl_r1} and \eqref{eq:anl_r2} in the large-$N$ limit,
instead of $r_i(\hat{t},N)$ evaluated at finite $N$ values.
This is because the difference between values at finite $N$ and the large-$N$ limit does not significantly affect the sample size estimate.
Therefore, we estimate the sample size as 
\begin{align}
    N_\mathrm{sample} \simeq
    \frac{\mathrm{Var}\qty(r_i(\hat{t},N))}{r_i(\hat{t})^2} \frac{1}{R^2}.
\end{align}

Table~\ref{tab:estm_nsmp} details the sample size required for the coefficients $r_1(\hat{t})$ and $r_2(\hat{t})$
with a relative statistical error of $R=1\%$ at $N=800$.
We also estimate the number of samples for $\hat{t}=3.0$ and $7.5$, along with $\hat{t}=6.0$.
The lower and upper bounds are estimated based on the two interpolation results of the variance.

For the subtracted coefficient $r_2(\hat{t})-r_1(\hat{t})^2$, 
the sample size required to achieve a relative error of $R=1\%$ increases, as
the mean value of the coefficient is quite small owing to large cancellations.
We roughly estimate the sample size required for the subtracted coefficient to achieve $R=1\%$ as follows.
The relative error is approximated by
\begin{align}
\dfrac{\delta\qty(r_2(\hat{t},N) - r_1(\hat{t},N)^2)}{r_2(\hat{t},N) - r_1(\hat{t},N)^2}
& \simeq
\dfrac{\delta r_2(\hat{t},N) + 2 r_1(\hat{t},N) \delta r_1(\hat{t},N)}{r_2(\hat{t},N) - r_1(\hat{t},N)^2}
\notag\\
& =
\dfrac{r_2(\hat{t},N) R_2 + 2 r_1(\hat{t},N)^2 R_1 }{r_2(\hat{t},N) - r_1(\hat{t},N)^2},
\end{align}
where $R_1$ and $R_2$ denote the relative errors of $r_1(\hat{t},N)$ and $r_2(\hat{t},N)$, respectively.
We can neglect the term with $R_1$ because the relative error is dominated by $R_2$
when the sample size increased, as indicated in Table~\ref{tab:estm_nsmp}.
The denominator $r_2(\hat{t},N) - r_1(\hat{t},N)^2$ is roughly constant at $\sim 0.008$,
and the two-loop coefficient $r_2(\hat{t},N)$ takes values $0.082\sim 0.095$
in the flow time region $[3.0,7.5]$, as depicted in Fig.~\ref{fig:r1_largeN_wide}.
Thus, the relative error of the subtracted coefficient is estimated as
\begin{align}
\dfrac{\delta\qty(r_2(\hat{t},N) - r_1(\hat{t},N)^2)}{r_2(\hat{t},N) - r_1(\hat{t},N)^2}
& \simeq
\dfrac{(0.082 \sim 0.095)}{0.008} R_2 \simeq (10\sim 12) R_2.
\end{align}
This implies that to achieve a 1\% relative error for the subtracted coefficient, we must reduce the relative error of $r_2$ to 0.1\%,
which requires approximately 100 times more samples than those required for $r_2$ to achieve the same 1\% relative error.

\begin{table}[ht]
    \tbl{Estimated sample sizes for coefficients  $r_1(\hat{t})$ and $r_2(\hat{t})$
    required to achieve a relative statistical error $R=1.0\%$ at $N=800$.}{
    \begin{tabular}{ccc}
    \toprule
    $\hat{t}$ & $N_\mathrm{sample}$ for $r_1(\hat{t})$ & $N_\mathrm{sample}$ for $r_2(\hat{t})$ 
     \\ \colrule
     $3.0$ &             $231$ $\sim$   $309$ &  7,027 $\sim$  9,238 \\
     $6.0$ & \phantom{1,}$814$ $\sim$ $1,253$ & 24,010 $\sim$ 38,894 \\
     $7.5$ &           $1,176$ $\sim$ $1,928$ & 33,754 $\sim$ 58,595 \\
     \botrule
     \end{tabular}
     \label{tab:estm_nsmp}}
\end{table}

\section{Summary and outlook}
\label{sec:summary_and_outlook}
In this study, we compute the gradient flow coupling expanded in terms of the lattice bare coupling for the TEK model in the large-$N$ limit using NSPT simulations.
By analyzing the flow time dependence of the perturbation coefficients, we identified a flow time region where 
the NSPT results are consistent with the analytical predictions,
taking into account both the discretization effect $\order{a^2/t}$ and the finite volume effect $\order{t^2/N}$.
At the one-loop level, we successfully extracted the coefficients $b_0$ and $f_1$ with $10\%$ accuracy using the current statistics.
However, large statistical errors prevented us from determining the coefficients beyond the two-loop level with precision.
To address this, we estimated the number of statistical samples required to accurately determine the two-loop beta function coefficient
using variance analysis and the large-$N$ factorization property of the model.

Finally, to give an outlook for the three-loop coefficient $r_3(\hat{t})$, we roughly estimate the sample size required using the same approach adopted for $\expval{r_2(\hat{t})}$, using Eq.~\eqref{eq:estm_nsmp}.
Our calculations indicate that at least $280, 000$ samples are necessary for $r_3(\hat{t},N)$ at $N=800$ and $\hat{t}=6$ to achieve a fixed relative error of $1\%$.
This is approximately $10$ times larger than the sample size required for $r_2(\hat{t})$, which ranges from $24,010\sim 38,894$ samples.
We also determined the computational time required for the gradient flow evolution at $N=800$ as $T_\mathrm{GF}=2,050$ [sec/conf]. 
Using single-node computation, we estimated that to achieve a $1\%$ relative error at $N=800$, 
determining $r_2(\hat{t})$ would require $570\sim 920$ days, while determining $r_3(\hat{t})$ would require approximately $\sim 6600$ days.
Thus, although the two-loop computation for $r_2$ seems feasible, determining the three-loop coefficient $r_3$
would be more challenging.
Furthermore, extracting the two-loop beta function coefficients from the flow time dependence is difficult,
as it requires a $0.1\%$ statistical error for $r_2(\hat{t})$.
We believe the same approach could be applied to other renormalization scheme couplings, such as
the Schr\"odinger functional scheme\cite{torrero2009_NSPT_SF},
the potential scheme, and the force scheme\cite{Horsley_2014}, at least at the one-loop level using NSPT in the large-$N$ limit.

\section*{Acknowledgments}
H.T. is supported by JST, the establishment of university fellowships toward the creation of science technology innovation, Grant Number JPMJFS2129.
K.-I.I. is supported in part by MEXT as ``Feasibility studies for the next-generation computing infrastructure''. 
M.O. is supported by JSPS KAKENHI Grant Number 21K03576.
The computations were carried out using the following computer resources;
 ITO subsystem-B offered under the category of general projects by the Research Institute for Information Technology, Kyushu University,
 Cygnus at the Center for Computational Sciences, University of Tsukuba, and
 SQUID at the Cybermedia Center, Osaka University under the support of the RCNP joint use program.

\appendix 

\section{Asymptotic behavior of the energy density at the tree-level}
\label{sec:tree_level_largen}
This appendix deals with the asymptotic behavior of the tree-level estimate of the energy density $\expval{\hat{t}^2E_W(\hat{t})/N}$ in the large-$N$ and large flow time limits.
The energy density defined with the plaquette is expressed in Eq.~\eqref{eq:energy_density_w},
while its analytic form\cite{perez2014suinfty} for the tree-level estimate is
\begin{align}
    \mathcal{N}_W(\hat{t}) \equiv &\left. \expval{\frac{\hat{t}^2E_W(\hat{t})}{N}} \right|_{\mathrm{tree}}
    = \frac{3\hat{t}^2}{2N^2} \sum_q' \mathrm{e}^{-2\hat{t}\hat{q}^2}
    \label{eq:lattice_ew}
\end{align}
where $\hat{q}^2=\sum_\mu \hat{q}_{\mu}^2,\ \hat{q}_\mu=2\sin(q_\mu/2)$ and $q_\mu = 2\pi m_\mu/\sqrt{N}\ (m_\mu=0,\dots,\sqrt{N}-1)$.
The primed summation means the summation excluding the zero mode $q=0$.
In the continuum theory, the tree-level coefficients is ${3}/{(128\pi^2)}$.
Eq.~\eqref{eq:lattice_ew} will become $3/(128\pi^2)$ in the large-$N$ limit at large flow time. 
The asymptotic form is derived as follows.

Equation~\eqref{eq:lattice_ew} can be rewritten as 
\begin{align}
    \mathcal{N}_W(\hat{t}) =
    \frac{3 \hat{t}^2}{2N^2} \mathrm{e}^{-16 \hat{t}}\left[\sum_{p=0}^{\sqrt{N}-1}\mathrm{e}^{4 \hat{t}\cos\qty(\frac{2\pi p}{\sqrt{N}})}\right]^4 - \frac{3 \hat{t}^2}{2N^2}.
\label{eq:series_bessel}
\end{align}
Using the generating function of the modified Bessel function of the first kind, we have:
\begin{align}
    \mathrm{e}^{\frac{z}{2}( \hat{t}+ \hat{t}^{-1})} =& 
    \sum_{n=-\infty}^{\infty}I_n(z) \hat{t}^n,
\label{eq:generating_bessel}
\end{align}
The term in the bracket of Eq.\eqref{eq:series_bessel} becomes
\begin{align}
    \sum_{p=0}^{\sqrt{N}-1} \mathrm{e}^{4\hat{t}\cos\qty(\frac{2\pi p}{\sqrt{N}})}
    =&
    \sum_{p=0}^{\sqrt{N}-1} \sum_{n=-\infty}^{\infty} I_n(4\hat{t}) \mathrm{e}^{i\frac{2\pi}{\sqrt{N}}pn}
    \\ =&
     \sqrt{N} \left[ I_0(4\hat{t}) + 2\sum_{k=1}^{\infty} I_{k\sqrt{N}}(4\hat{t}) \right].
\end{align}
Substituting the above equation into Eq.~\eqref{eq:series_bessel} yields
\begin{align}
    \mathcal{N}_W(\hat{t}) =&
    \frac{3\hat{t}^2}{2N^2} \qty(\sqrt{N})^4 \mathrm{e}^{-16\hat{t}} \left[ I_0(4\hat{t}) + 2\sum_{k=1}^{\infty} I_{k\sqrt{N}}(4\hat{t}) \right]^4
    - \frac{3\hat{t}^2}{2N^2}.
    \label{eq:bessel_form}
\end{align}

For the asymptotic form in the large-$N$ limit, we use the asymptotic form of the Bessel function
for large orders at a fixed $z$:
\begin{align}
    I_\nu(z)\sim \frac{1}{\sqrt{2\pi\nu}}\left(\frac{\mathrm{e}z}{2\nu}\right)^\nu.
\end{align}
In the large-$N$ limit, Eq.~\eqref{eq:bessel_form} becomes
\begin{align}
    \mathcal{N}_W(\hat{t}) \sim &
    \frac{3\hat{t}^2}{2N^2} \qty(\sqrt{N})^4 \mathrm{e}^{-16\hat{t}} \left[ I_0(4\hat{t}) + 
    2\sum_{k=1}^\infty \frac{1}{\sqrt{2\pi k\sqrt{N}}}\left(\frac{2\mathrm{e}\hat{t}}{k\sqrt{N}}\right)^{k\sqrt{N}} \right]^4
    - \frac{3\hat{t}^2}{2N^2}.
\end{align}
The second term in the brakets denotes a nonanalytic dependence on $N$,
which complicates the $N$ dependence in the mid-flow time region.
In the large-$N$ limit, only the first term, $I_0(4\hat{t})$, contributes to the energy density:
\begin{align}
    \mathcal{N}_W(\hat{t}) =&
    \frac{3\hat{t}^2}{2}\mathrm{e}^{-16\hat{t}} I_0(4\hat{t})^4  \quad (N\to\infty).
\label{eq:tree_largen}
\end{align}

Substituting the asymptotic form of the Bessel function for a large argument $z$ into Eq.~\eqref{eq:bessel_form} 
yields the following asymptotic form for the energy density:
\begin{align}
    \mathcal{N}_W(\hat{t}) 
    \sim&
    \frac{3}{128\pi^2}
    \left[1 + \frac{1}{8\hat{t}}+\cdots \right]
    - \frac{3\hat{t}^2}{2N^2},
\label{eq:asympt_ew}
\end{align}
where we omit nonanalytic terms originating from higher-order Bessel functions.
This is the asymptotic behavior in the large-$N$ limit and large flow time regions.
The leading term is consistent with the continuum result $3/(128\pi^2)$,
and the leading finite-$N$ correction term is the last term $-{3\hat{t}^2}/{2N^2}$. 
The effect of the lattice artifact $\order{a^2/t}$ manifests as the second term in the bracket.

\begin{figure}[t]  
    \centering
        \includegraphics[clip,scale=\figscale]{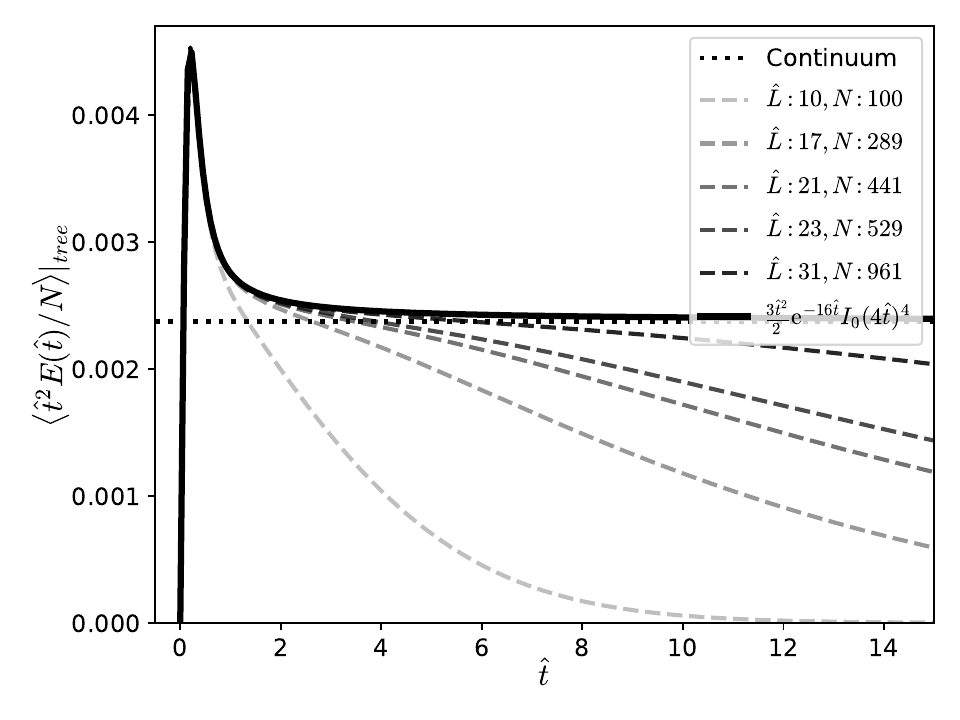}
        \caption{Flow time dependence of energy density $\expval{\hat{t}^2 E(\hat{t})/N}$ and continuum result $3/128\pi^2$.
        The dashed curves represent the finite $N$ results derived from Eq.~\eqref{eq:lattice_ew}.
        }
    \label{fig:tree_all}
\end{figure}

Figure~\ref{fig:tree_all} the flow time dependence of the energy density $\expval{\hat{t}^2E_W(\hat{t})/N}$ at finite-$N$.
The solid curve represents the energy density at infinite $N$,while the dotted line represents the continuum result $3/128\pi^2$.
The broken curves denote the finite-$N$ results derived from Eq.~\eqref{eq:lattice_ew} for five matrix sizes.
In the mid-flow time region, the finite-$N$ corrections appear complicated owing 
to the higher-order Bessel functions in Eq.~\eqref{eq:bessel_form}. 

\renewcommand{\figscale}{0.38}
\begin{figure}[t]  
    \centering
        \includegraphics[clip,scale=\figscale]{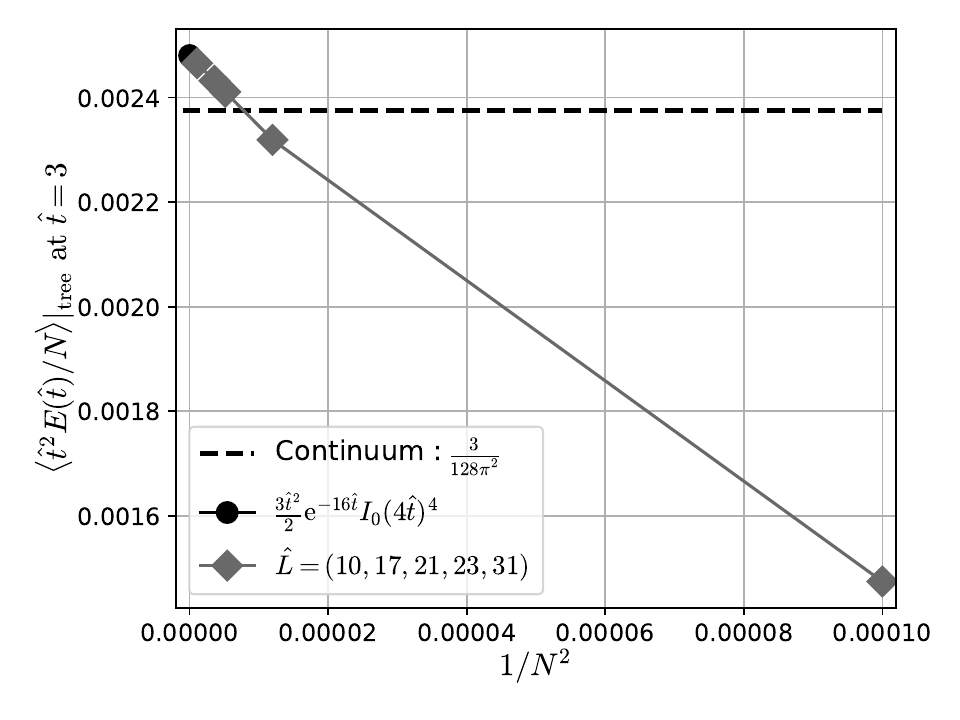}\hfill
        \includegraphics[clip,scale=\figscale]{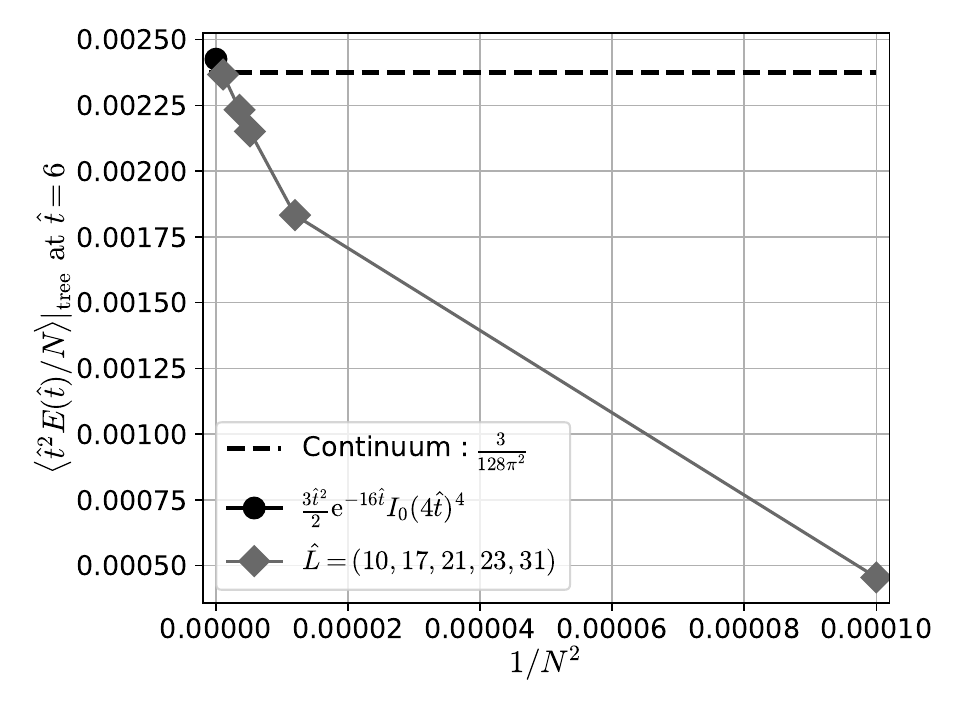}
        \includegraphics[clip,scale=\figscale]{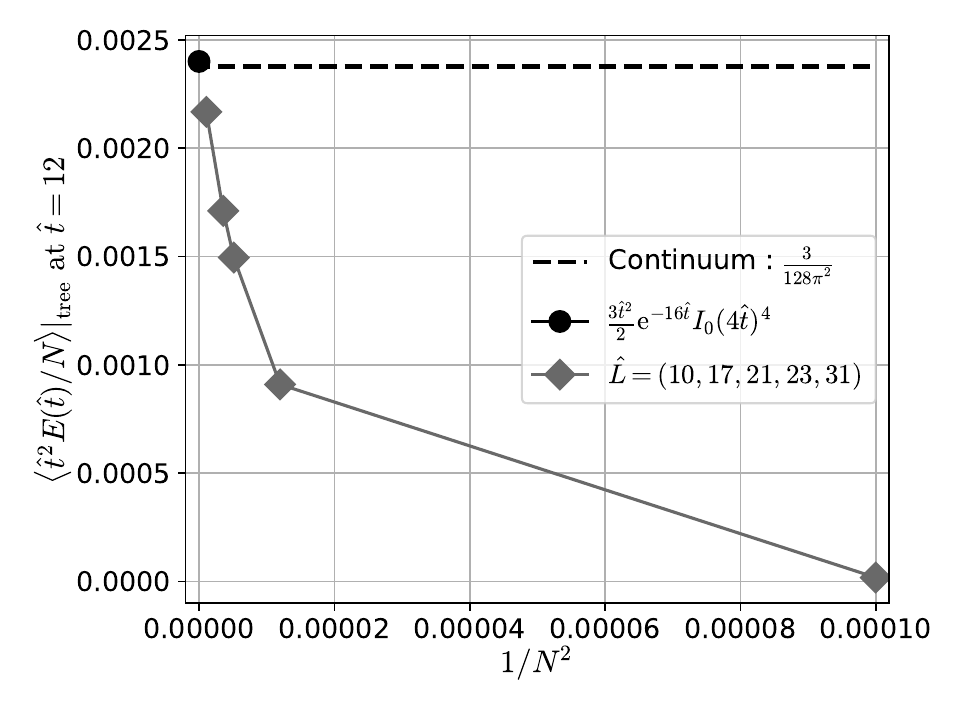}\hfill
        \includegraphics[clip,scale=\figscale]{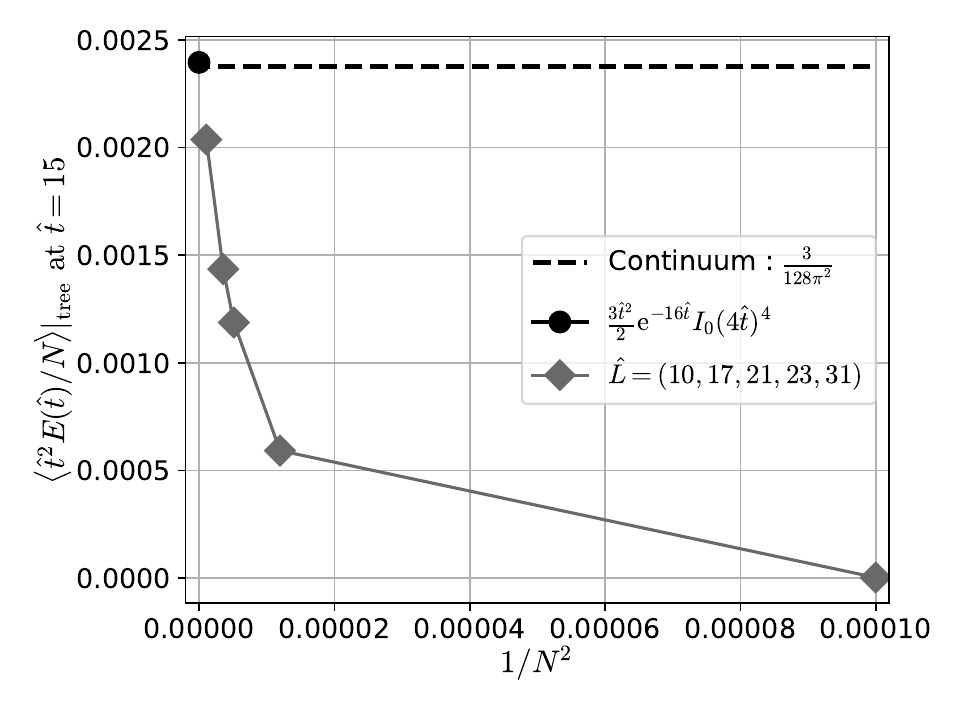}
        \caption{ $N$ dependence of the tree-level energy density at $\hat{t}=3,6,12,15$.
        }
        \label{fig:tree_vs_vol}
\end{figure}

Figure~\ref{fig:tree_vs_vol} illustrates the $N$ dependence of the energy density 
at four fixed time slices ($\hat{t}=3,6,12,15$).
For small flow times, $\hat{t}=3$ and $6$, the $N$-dependence is almost linear in $1/N^2$.
However, a lattice cut-off effect $\order{a^2/\hat{t}}$ emerges in the large-$N$ limit.
At large flow times, such lattice cut-off effects are suppressed with $\order{a^2/\hat{t}}$,
and the large-$N$ results are consistent with the continuum ones.
However, the linearity in $1/N^2$ is not guaranteed for small $N$ owing to
the finite volume effects of the higher order Bessel functions in Eq.~\eqref{eq:bessel_form}. 
A linear extrapolation in $1/N^2$ using smaller $N$ for higher order coefficients would miss the correct large $N$ limit in the large flow time region.

\section{Asymptotic behavior of the variance of the energy density at the tree level}
\label{sec:var_ene}
This section delves into the asymptotic behavior of the variance of 
the energy density $\hat{t}^2E_W(\hat{t})/N$ defined using the Wilson plaquette.
The variance of the energy density $\hat{t}^2E_W(\hat{t})/N$ is expressed as 
\begin{align}
    \mathrm{Var}\qty(\dfrac{\hat{t}^2E_W(\hat{t})}{N}) \equiv \dfrac{1}{N^2}\qty(\expval{\hat{t}^4E_W(\hat{t})^2} - \expval{\hat{t}^2E_W(\hat{t})}^2).
\end{align}
The analytic form for the tree-level estimate of the variance $\mathrm{Var}(\hat{t}^2E_W(\hat{t}))_\mathrm{tree}$ is 
\begin{align}
    \mathrm{Var}\left.\qty(\dfrac{\hat{t}^2E_W(\hat{t})}{N})\right|_\mathrm{tree} = \frac{3\hat{t}^4}{2N^4}\sum_p' \mathrm{e}^{-4\hat{t}\hat{p}^2}.
\label{eq:ana_var_ew}
\end{align}

Using the same analysis approach adopted in \ref{sec:tree_level_largen},
the variance can be expressed in terms of the modified Bessel function of the first kind:
\begin{align}
    \mathrm{Var}\left.\qty(\dfrac{\hat{t}^2E_W(\hat{t})}{N})\right|_\mathrm{tree} =&
    \frac{3\hat{t}^4}{2N^2}e^{-32\hat{t}} \qty(I_0(8\hat{t}) + 2\sum_{s=1}^\infty I_{s\sqrt{N}}(8\hat{t}))^4 - \frac{3\hat{t}^4}{2N^4}.
\end{align}
In the large-$N$ limit and large $\hat{t}$ regions, the variance behaves as 
\begin{align}
    \mathrm{Var}\left.\qty(\dfrac{\hat{t}^2E_W(\hat{t})}{N})\right|_\mathrm{tree} 
 \sim
    \frac{3\hat{t}^2}{2N^2}\qty[ \frac{1}{(16\pi)^2}\qty(1 + \frac{1}{16\hat{t}}+\cdots) - \frac{\hat{t}^2}{N^2}].
\label{eq:asympt_var_ew}
\end{align}
The asymptotic behavior of the variance is similar to that of the tree-level estimate of the mean value in Eq.~\eqref{eq:asympt_ew}.
At large flow times with small $N$ values, the finite volume correction $\order{\hat{t}^2/N^2}$
in the bracket of Eq.~\eqref{eq:asympt_var_ew} becomes significant.
Therefore, we could miss the zero variance in the large-$N$ limit if we relied
on a simple linear extrapolation in $1/N^2$ from small $N$ values to derive higher-order coefficients, as illustrated in Fig.~\ref{fig:variance_t6}.

\section{Numerical results at finite \texorpdfstring{$N$}{N}}
\label{sec:some_result}
Tables~\ref{tab:finite_r1}, \ref{tab:finite_r2}, and \ref{tab:finite_r3} present the results of the coefficients 
$r_1(\hat{t},N)$, $r_2(\hat{t},N)$, and $r_3(\hat{t},N)$ derived using the clover-type energy density at finite $N$ values of $289$, $441$, and $529$,
along with the corresponding large-$N$ extrapolation results at several flow times.

\begin{table}[ht]
    \tbl{Results of the coefficient $r_1(\hat{t})$ at finite $N$ values of $289$, $441$, and $529$,
    along with the extrapolation results at certain flow times.}{
    \begin{tabular}{clllll}
    \toprule
     \multicolumn{1}{c}{Flow time}  & \multicolumn{1}{c}{$N=289$} & \multicolumn{1}{c}{$N=441$} & \multicolumn{1}{c}{$N=529$} & \multicolumn{1}{c}{Extrapolation} & \multicolumn{1}{c}{$\chi^2/N_\mathrm{dof}$}\\
     \hline
0.60 & 0.2438( 3) & 0.2440( 2) & 0.2440( 2) & 0.2441( 3) & 0.936 \\
1.20 & 0.2568( 5) & 0.2567( 5) & 0.2568( 5) & 0.2567( 6) & 0.953 \\
1.80 & 0.2652( 8) & 0.2645( 8) & 0.2647( 7) & 0.2643( 9) & 0.992 \\
2.40 & 0.2715(10) & 0.2703(10) & 0.2705(10) & 0.2699(13) & 1.005 \\
3.00 & 0.2767(13) & 0.2751(13) & 0.2753(12) & 0.2745(16) & 1.004 \\
3.60 & 0.2812(15) & 0.2793(16) & 0.2794(15) & 0.2783(19) & 1.002 \\
4.20 & 0.2852(17) & 0.2829(18) & 0.2830(17) & 0.2817(22) & 1.002 \\
4.80 & 0.2890(19) & 0.2863(21) & 0.2862(19) & 0.2848(24) & 1.004 \\
5.40 & 0.2924(21) & 0.2895(23) & 0.2892(21) & 0.2876(27) & 1.007 \\
6.00 & 0.2956(23) & 0.2924(25) & 0.2921(24) & 0.2904(29) & 1.011 \\
6.60 & 0.2986(25) & 0.2953(27) & 0.2948(26) & 0.2930(32) & 1.015 \\
7.20 & 0.3013(27) & 0.2980(29) & 0.2974(28) & 0.2957(34) & 1.020 \\
     \botrule
     \end{tabular}
     \label{tab:finite_r1}}
\end{table}

\begin{table}[ht]
    \tbl{Same as Table~\ref{tab:finite_r1}, but for coefficient $r_2(\hat{t},N)$.}{
    \begin{tabular}{clllll}
    \toprule
     \multicolumn{1}{c}{Flow time}  & \multicolumn{1}{c}{$N=289$} & \multicolumn{1}{c}{$N=441$} & \multicolumn{1}{c}{$N=529$} & \multicolumn{1}{c}{Extrapolation} & \multicolumn{1}{c}{$\chi^2/N_\mathrm{dof}$} \\
     \hline
0.60 & 0.0660( 4) & 0.0655( 4) & 0.0659( 4) & 0.0656( 5) & 1.746 \\
1.20 & 0.0731( 8) & 0.0727( 8) & 0.0734( 8) & 0.0731(10) & 1.457 \\
1.80 & 0.0775(13) & 0.0771(12) & 0.0781(13) & 0.0779(16) & 1.439 \\
2.40 & 0.0808(18) & 0.0802(17) & 0.0816(17) & 0.0811(22) & 1.493 \\
3.00 & 0.0835(22) & 0.0824(22) & 0.0843(22) & 0.0835(27) & 1.559 \\
3.60 & 0.0860(26) & 0.0840(27) & 0.0865(27) & 0.0851(33) & 1.613 \\
4.20 & 0.0884(30) & 0.0854(31) & 0.0884(32) & 0.0864(38) & 1.651 \\
4.80 & 0.0906(33) & 0.0866(36) & 0.0900(36) & 0.0874(43) & 1.676 \\
5.40 & 0.0927(37) & 0.0877(40) & 0.0914(40) & 0.0883(48) & 1.693 \\
6.00 & 0.0946(40) & 0.0887(44) & 0.0928(44) & 0.0891(52) & 1.706 \\
6.60 & 0.0965(43) & 0.0896(48) & 0.0941(48) & 0.0899(57) & 1.717 \\
7.20 & 0.0982(46) & 0.0905(52) & 0.0954(52) & 0.0906(61) & 1.728 \\
     \botrule
     \end{tabular}
     \label{tab:finite_r2}}
\end{table}

\begin{table}[ht]
    \tbl{Same as Table~\ref{tab:finite_r1}, but for coefficient $r_3(\hat{t},N)$.}{
    \begin{tabular}{clllll}
    \toprule
     \multicolumn{1}{c}{Flow time}  & \multicolumn{1}{c}{$N=289$} & \multicolumn{1}{c}{$N=441$} & \multicolumn{1}{c}{$N=529$} & \multicolumn{1}{c}{Extrapolation} & \multicolumn{1}{c}{$\chi^2/N_\mathrm{dof}$} \\
     \hline
0.60 & 0.0173( 7) & 0.0190( 6) & 0.0181( 7) & 0.0191( 9) & 2.574 \\
1.20 & 0.0197(15) & 0.0220(14) & 0.0202(15) & 0.0217(19) & 2.142 \\
1.80 & 0.0217(24) & 0.0248(23) & 0.0213(24) & 0.0235(30) & 2.408 \\
2.40 & 0.0234(32) & 0.0278(32) & 0.0225(33) & 0.0256(41) & 2.612 \\
3.00 & 0.0248(41) & 0.0307(41) & 0.0239(43) & 0.0280(53) & 2.669 \\
3.60 & 0.0262(49) & 0.0336(50) & 0.0255(52) & 0.0306(63) & 2.622 \\
4.20 & 0.0274(57) & 0.0362(58) & 0.0272(61) & 0.0332(74) & 2.525 \\
4.80 & 0.0285(65) & 0.0387(66) & 0.0289(70) & 0.0359(83) & 2.410 \\
5.40 & 0.0296(72) & 0.0410(74) & 0.0307(79) & 0.0385(92) & 2.298 \\
6.00 & 0.0306(78) & 0.0432(81) & 0.0325(87) & 0.0411(101) & 2.194 \\
6.60 & 0.0315(84) & 0.0452(88) & 0.0343(95) & 0.0436(109) & 2.102 \\
7.20 & 0.0323(90) & 0.0471(95) & 0.0360(103) & 0.0460(116) & 2.022 \\
     \botrule
     \end{tabular}
     \label{tab:finite_r3}}
\end{table}

\bibliographystyle{ws-ijmpa}
\bibliography{biblio}

\end{document}